\documentclass[preprint,aps,amsmath,nofootinbib,tightenlines]{revtex4}
\usepackage{graphicx}
\usepackage{bm}
\usepackage{epsfig}
\usepackage{graphics}
\usepackage{amsfonts,amssymb}

\def\xslash#1{{\rlap{$#1$}/}}
\def\Dsl{\hbox{/\kern-.6000em D}} %roman D

\def\dsl{\,\raise.15ex\hbox{/}\mkern-13.5mu D}
\def\bsigma{\mbox{\boldmath $\sigma$}}

\def\bsigma{\mbox{\boldmath $\sigma$}}

\def\abs#1{\left| #1 \right|}
\def\ltap{\ \raise.3ex\hbox{$<$\kern-.75em\lower1ex\hbox{$\sim$}}\ }
\def\gtap{\ \raise.3ex\hbox{$>$\kern-.75em\lower1ex\hbox{$\sim$}}\ }
\def\OMIT#1{}

\def\lsim{\mathrel{\raise.3ex\hbox{$<$\kern-.75em\lower1ex\hbox{$\sim$}}}}
\def\gsim{\mathrel{\raise.3ex\hbox{$>$\kern-.75em\lower1ex\hbox{$\sim$}}}}

\newcommand{\nn}{\nonumber}

\newcommand{\me}[3]{\ensuremath{\left\langle{#1}\vphantom{#2 #3}
\right|{#2}\left|\vphantom{#1 #2}{#3}\right\rangle}}

\newcommand{\bmk}{\mathbf k}
\newcommand{\bmp}{\mathbf p}

\newcommand{\bmq}{\mathbf q}

\newcommand{\bmone}{\mathbf 1}

\newcommand{\bmsigma}{\mathbf \bsigma}

\newcommand{\bmx}{\mathbf x}

\catcode`\@=11
\def\slash{\mathpalette\make@slash}
\def\make@slash#1#2{\setbox\z@\hbox{$#1#2$}%
  \hbox to 0pt{\hss$#1/$\hss\kern-\wd0}\box0}
\catcode`\@=12 % @ signs are no longer letters
\begin{document}
%%%%%%%%%%%%%%%%%%%%%%%%%%%%%%%%%%%%%%%%%%
%Some more stuff to get graphics to work
%\ifpdf
%\DeclareGraphicsExtensions{.pdf, .jpg}
%\else
%\DeclareGraphicsExtensions{.eps, .jpg}
%\fi
%%%%%%%%%%%%%%%%%%%%%%%%%%%%%%%%%%%%%%%%%%

%%%%%%%%%%%%%%%%%%%%%%%%%%%%%%%%%%%%%%%%%%
%Define Title, Author, Address, Preprint#

\preprint{ \vbox{ \hbox{MPP-2006-46} 
%\hbox{hep-ph/yymmnnn}  
}}

\title{\phantom{x}\vspace{0.5cm} 
The soft-energy region in the radiative decay \\[3mm]
of bound states
\vspace{1.0cm} }

\author{Pedro~D.~Ruiz-Femen\'\i a \vspace{0.5cm}}
\affiliation{Max-Planck-Institut f\"ur Physik \\
(Werner-Heisenberg-Institut) \\
F\"ohringer Ring 6,
80805 M\"unchen, Germany\vspace{1cm}
\footnote{Electronic address: ruizfeme@mppmu.mpg.de}\vspace{1cm}}

%\date{\today\\ \vspace{1cm} }

%%%%%%%%%%%%%%%%%%%%%%%%%%%%%%%%%%%%%%%%%%
\begin{abstract}
\vspace{0.5cm}
\setlength\baselineskip{18pt}
The orthopositronium decay to three photons is studied in the phase-space region where one of the photons has an 
energy comparable to the relative three-momentum of the $e^+e^-$ system ($\omega\sim m\alpha$). 
The NRQED computation in this regime
shows that the dominant contribution arises from distances $\sim 1/\sqrt{m \omega}$, which allows 
to treat the Coulomb interaction perturbatively. 
The small-photon energy expansion of the 1-loop decay spectrum from full QED yields the same result
as the effective theory. By doing the threshold expansion of the 1-loop QED amplitude we 
confirm that the leading term arises from a loop-momentum region where 
$q^0\sim \mathbf{q}^2/m \sim \omega$. This corresponds to a new non-relativistic loop-momentum region, which has to
be taken into account for the description of a non-relativistic particle-antiparticle system
that decays through soft photon emission.
\end{abstract}
% \pacs{12.39.Hg,11.10.St,12.38.Bx}
\maketitle

%%%%%%%%%%%%%%%%%%%%%%%%%%%%%%%%%%%%%%%%%%

% \tighten
\newpage
%%%%%%%%%%%%%%%%%%%%%%%%%%%%%%%%%%%%%%%%%%
%Main body of the paper
%\setlength\baselineskip{15pt}

%
%
%
\section{Introduction}
\label{sec:intro}
Effective field theories (EFT's) and asymptotic expansions have become standard tools in the description
of the radiative decay of bound states. Their success lies on their ability to select
the relevant dynamics in the different kinematic regions defined by the relative size of the 
radiated photon energy 
as compared to the bound state mass. Methods based on the use of operator product expansions
and effective Lagrangians 
do not only provide more efficient ways to perform the computations but also allow to
extract physical interpretations which may go unnoticed in a full theory approach.

Apart from being of physical relevance in itself, the Positronium 
system provides a testing ground for the EFT concepts and techniques devised 
in the study of radiative decay amplitudes, that could eventually be applied to
the description of the more intricate radiative decays of quarkonia.
In this paper we illustrate which are the characteristic features of the
radiative decay spectrum when the energy of the photon lies in
the soft-energy region ($\omega\sim m\alpha$) through the study of
the $3\gamma$ annihilation amplitude of 
the Positronium spin triplet ground state (orthopositronium: o-Ps). The soft-energy 
region is accessible in the o-Ps$\to 3\gamma$ decay because one of the final state photons can 
have an arbitrarily small energy, the other two being hard photons with
energy $\lsim m$. 
The process can be then viewed as the radiative version of the o-Ps$\to 2\gamma$ decay.

In a previous paper~\cite{our} the o-Ps differential decay spectrum was calculated
in the region where the photon energy is comparable to the Positronium binding energy,
$\omega\sim m\alpha^2$. Binding effects were included for the first time in the Ps
structure function
using non-relativistic QED (NRQED)~\cite{NRQED}, 
and it was found that they are essential 
to reach agreement with the Low's theorem prediction for the $\omega\to 0$ behaviour
of this decay~\cite{our,oPsproc}. Only the leading term in the multipole
expansion of the radiated photon field (dipole approximation)
was required for the calculation,
in accordance with the NRQED velocity counting rules for photons with $\omega\sim m\alpha^2$~\cite{powercounting}.
Based on general considerations about the power-counting of higher-order multipoles in
the non-relativistic description of the o-Ps decay process,
it was then argued by Voloshin~\cite{voloshin} that the dipole approximation used for
photon energies $\omega\sim m\alpha^2$
should also hold for a description of the o-Ps photon spectrum in the whole range
$\omega\ll m$, thus enlarging the validity region of the formula given in
Ref.~\cite{our}. In particular, it was shown by Voloshin that the 
expansion of the above-mentioned formula in the region $m\alpha^2\ll \omega \ll m$ 
is actually a series in $\alpha\sqrt{m/\omega}\sim\sqrt{\alpha}$ rather than in 
integer powers of $\alpha$. 

The origin of such unnatural expansion has to be traced back to 
the fact that the main contribution in the EFT calculation arises from
distances in the $e^+e^-$ system of order $r~\sim 1/\sqrt{m\omega}$, that are
much smaller than the size of the Ps atom $r\sim 1/m\alpha$. Translated into momentum
space, this implies that there is a kinematic region where the relative 3-momentum
of the non-relativistic $e^+e^-$ pair obeys the non-relativistic relation $p^0\sim\bmp^2/m\sim\omega$. 
This scaling does not correspond to any of the known modes that have 
been identified in the common applications of non-relativistic EFT's, and
should be compared to the scaling of 
the familiar potential modes, $p^0\sim \bmp^2/m \sim m\alpha^2$, characteristic of the heavy particles that form the
bound state, and that of the soft modes, $p^0\sim\omega\gg\bmp^2/m$.

In view of these novel features,
it is worth questioning if the non-relativistic EFT framework devised
to describe the o-Ps decay for $\omega\sim m\alpha^2$ in Ref.~\cite{our}
also provides the proper expansion when extended 
to the whole $\omega\ll m$ region, as advocated by Voloshin. 
It is the main purpose of this work to confirm that this is indeed the case by
comparing the analytic results for the EFT and full QED spectrum in
the $\omega\sim m\alpha$ region,
where, as we argue, the usual perturbative QED expansion can already be applied. 
Such comparison has become feasible after a recent evaluation of 
the 1-loop QED o-Ps$\to 3\gamma$ phase-space distribution~\cite{adkins}, that we use
to determine the QED spectrum for soft photon energies. 
The agreement between both computations shows that the soft-energy region
provides a regime where the EFT and the perturbative QED calculations can have
a smooth matching. 
Moreover, we shall confirm by explicitly computing the o-Ps$\to 3\gamma$ decay spectrum
in NRQED beyond the dipole approximation that the $(\omega/m)^k$ suppression
advocated by Voloshin for the higher-order multipoles  
is of application in the soft-energy region.

It is also the aim of this work to show that the 
momentum scale $p_0\sim \bmp^2/m \sim \omega$ that rules the
behaviour of the corrections to the NRQED o-Ps decay amplitude, corresponds to
a new loop-momentum region that has to be 
taken into account for a successful application of the
threshold expansion method~\cite{beneke} to QED and QCD
loop diagrams
involving a particle-antiparticle system decaying through 
the emission of a soft photon. In the conventional loop expansion,
soft photon radiation from a heavy particle-antiparticle system introduces
a soft energy component $\omega\sim m\alpha$ in the zero component of the heavy particle momenta.
Non-relativistic poles in
massive propagators of the form $(p^0-\bmp^2/2m-\omega)^{-1}$ can be thus found in
these loops, giving rise to a contribution from the loop-momentum region 
$p_0\sim \bmp^2/m \sim \omega$.
It is a well-known fact that momentum regions that have not been considered 
previously may become relevant for some kinematic configurations
or at higher loop order in the asymptotic expansion of integrals based on the 
``method of regions"~\cite{smirnov}, 
and looking for missing regions is a required task in order to check the robustness
of the method.
Likewise, it is of conceptual
importance to understand the role of this new momentum scale
in the EFT frameworks that describe the radiative decays of 
heavy particle-antiparticle systems 
in the soft-energy region, which we discuss in this paper.

The outline of the paper is as follows. In Sec.~\ref{sec:NRQED} we detail the EFT calculation
of the o-Ps$\to 3\gamma$ decay amplitude without multipole expanding the electromagnetic interaction.
We also show how relativistic corrections that contribute
to the decay amplitude in the whole photon energy range $\omega\ll m$ can be accommodated 
in the EFT framework. In particular, the recoil of the intermediate virtual $e^+e^-$ pair,
which yields also an $\omega/m$ correction, is included in our calculation.
In Sec.~\ref{sec:QED} we perform the $\omega\to 0$ limit of the 1-loop QED spectrum directly from 
the 1-loop phase-space distribution computed recently by Adkins~\cite{adkins}. The 1-loop QED
amplitude in this regime is also obtained with the asymptotic expansion of the graphs near threshold in Sec.~\ref{sec:threshold}.
A discussion on the EFT description of the new loop-momentum region found is postponed to the
end of the latter section. Finally in Sec.~\ref{sec:quarkonium} we suggest the use of the threshold 
expansion within the NRQCD factorization framework to compute the short-distance coefficients 
in quarkonium radiative decays at soft energies. 
The Appendix~\ref{sec:appen1}
collects the exact formulas for the multipole calculation of the NRQED spectrum, while
Appendix~\ref{sec:appen2} shows another example of the asymptotic expansion method
applied to a radiative amplitude.

\section{NRQED computation of the decay spectrum for $\omega \ll m$} 
\label{sec:NRQED}

\begin{figure}
\begin{center}
\includegraphics[width=9cm]{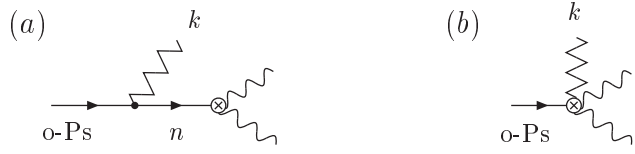}
\end{center}
\caption{NRQED graphs for the o-Ps$\to 3\gamma$ decay. The allowed intermediate states $n$
correspond to those of the spectrum of $H_0$. The zig-zag line represents the low-energy photon. 
The black dot in $(a)$ denotes either a $\bmp\cdot\mathbf{A}$ or a $\bmsigma\cdot \mathbf{B}$ interaction. 
Graph $(b)$ represents the radiation of the low-energy photon directly from the $2\gamma$ annihilation vertex.\label{fig:NRQEDgraphs}}
\end{figure}

The NRQED framework used in Ref.~\cite{our} provides a systematic way to compute 
bound state effects in the the o-Ps decay spectrum when the
photon energy is much smaller than the electron mass.
Contrary to the usual relativistic approach where the 3-photon annihilation is considered to take
place at very short distances as compared to the range of the electromagnetic binding force between $e^+e^-$,
the non-relativistic description takes into account that there is a long-distance part 
in the o-Ps decay process when one the photons in the final state is not
hard ($\omega\ll  m$). In the latter case, the decay proceeds in two steps: 
the low-energy photon is first radiated from the bound state, making a
transition from the C-odd ground state o-Ps (${}^3S_1$) to a C-even Positronium state,
which subsequently decays into two photons (see the EFT graph of Fig.~\ref{fig:NRQEDgraphs}a). 
The decay amplitude of the intermediate C-even 
$e^+e^-$ state has no long-distance contribution, since both photons must be hard. 
The emission of the low energy photon from o-Ps is
described by the Coulomb Hamiltonian of the $e^+e^-$ system 
in interaction with a quantized electromagnetic field:
\begin{eqnarray}
H &=& H_0+H_{\text{int}} \,, \nn\\[3mm]
H_0&=&\frac{\mathbf{P}^2}{4m}+H_C \;\;,\;\; H_C = \frac{\mathbf{p}^2}{m} -\frac{\alpha}{r} \,, \nn\\[2mm]H_{\text{int}} &=& - { e \over m}\, \mathbf{p}_1\cdot \mathbf{A}(\mathbf{x}_1)
+{ e \over m}\, \mathbf{p}_2\cdot \mathbf{A}(\mathbf{x}_2)
-\mu\;\bm{\sigma}_{\phi}\cdot\mathbf{B}(\mathbf{x}_1)-
\mu\;\bm{\sigma}_{\chi}\cdot\mathbf{B}(\mathbf{x}_2) \,,
\label{1.1}
\end{eqnarray}
where we have used the center of mass variables ($r\equiv |\mathbf{x}|$)
$$
\mathbf{p}=(\mathbf{p}_1-\mathbf{p}_2)/2 \ \ , \ \
\mathbf{x}=\mathbf{x}_1-\mathbf{x}_2 \ \ , \ \, 
\mathbf{P}=\mathbf{p}_1+\mathbf{p}_2 \ \ , \ \, 
\mathbf{X}=\frac{\mathbf{x}_1+\mathbf{x}_2}{2}
\,,
$$
with the subindices ${1,2}$ referring to the electron and the positron, respectively, and $\bm{\sigma}_{\phi},\,\bm{\sigma}_{\chi}$ 
being the Pauli matrices acting on the electron and positron spinors ($\mu=e/2m$). 
The terms shown in the Hamiltonian $H_{\text{int}}$ are the leading ones in the non-relativistic expansion. 
Relativistic effects can be included through higher dimensional operators suppressed by powers of $1/m$ 
(see \textit{e.g.}~\cite{manohar}), but will not be needed for the purposes of this work. 

Since the computation of the photon spectrum in Ref.~\cite{our} was intended
for photon energies comparable to the Positronium binding energy, $\omega\sim m\alpha^2$, 
the interaction Hamiltonian was used in the dipole approximation limit. This approximation
amounts to evaluating the vector potential $\mathbf{A}$ in the center of mass of the 
Positronium system ({\it i.e.} at $\bmx=0$), which is fully legitimate for radiated photons with wavelengths
much larger than the characteristic size of the Positronium atom ($a=2a_0=2/m\alpha$). 
Higher order multipoles arise as a Taylor series in the relative coordinate $\mathbf{x}$ and 
are suppressed by powers of $\omega/a$ under the assumption that the relevant amplitudes 
one has to compute involve integrations to spatial extents of order $\sim a$. 
For photons with larger energies,
$\omega\gg m\alpha^2$, this premise will invalidate the use of the
multipole expansion. 

However, in the case of the o-Ps system that undergoes a radiative transition
before decaying, and as it has been properly pointed out by Voloshin~\cite{voloshin}, 
the scale $(m\alpha)^{-1}$ does not constraint the maximum photon energy for which we can
apply the multipole expansion. The reason is that after the soft photon is radiated we have to
consider all possible $e^+e^-$ intermediate states with the right quantum numbers. The 
propagation of these $e^+e^-$ states is described by the Green's function obtained from 
the Hamiltonian $H_C$,
\begin{eqnarray}
\left( H_C + {\kappa^2 \over m} \right)   G\left( \mathbf{x}, \mathbf{y}, \kappa \right) = \delta\left( \mathbf{x} - \mathbf{y} \right)\,,
\label{1.2}
\end{eqnarray}
at energy 
\begin{eqnarray}
-\kappa^2/m=E_0-\omega-\omega^2/4m\,,
\label{1.3}
\end{eqnarray}
(see Eq.~(\ref{1.8}) below). Since the annihilation of the 
intermediate $e^+e^-$ pair into two photons takes place at small distances,
the amplitude for the full process is thus given by a convolution of the o-Ps
ground state and the Green's function. In coordinate space, the Green's function
has a characteristic size ruled by the exponential factor $\exp(-\kappa r)$. 
We note that for soft photons of energy $m\alpha^2\ll \omega \ll m$, this exponential 
factor constrains the relevant integration region to distances of order 
$\kappa^{-1}\sim  \sqrt{m \omega}$, much smaller than the spatial extent
of the initial Positronium atom. Therefore, the characteristic distance that
enters in the multipole expansion of the soft-photon emission in
the o-Ps$\to 3\gamma$ decay amplitude is determined by the 
falloff of the intermediate Green's function rather than by the size of
the Ps atom, so the series of multipoles is actually 
an expansion in powers of $\omega r\sim \sqrt{\omega/m}$ which 
can be used as long as $\omega\ll m$. Note also that for the intermediate
pair, an iterative computation of the Coulomb Green's function,
$G=G^{f}+G^{f} V_C G^{f}+\dots$,
with $G^{f}$ the free Green's function,
shows that adding a Coulomb exchange generates a term 
$$\int d^3\mathbf{x} \,V_C\,G^{f} \sim {\alpha m\over \kappa}\sim \alpha\sqrt{{m \over \omega}}\ll 1\,,
\quad \mbox{for}\quad m\alpha^2\ll \omega\ll m\,,
$$
as $G^{f}\sim m\exp(-\kappa r)/r$.
Therefore the Coulomb interaction can be treated perturbatively in the 
virtual $e^+e^-$ system after soft photon radiation.

It is illustrative to show how the latter is realized 
in the actual computation of the o-Ps
decay spectrum without multipole expanding the electromagnetic potential
of the radiated photon. %(i.e. without dropping the photon wavefunctions)
For this purpose let us now consider the NRQED calculation 
of Ref.~\cite{our} and keep the full $\mathbf{x}_i$ dependence of the various
terms in the Hamiltonian~(\ref{1.1}). We start with the $\mathbf{p}\cdot\mathbf{A}$
electric amplitude\footnote{We refer to the amplitudes as of `electric' type
if there is a change in the parity between the atomic states (of `magnetic' type otherwise).}.
Using time-ordered perturbation theory (TOPT),
the amplitude corresponding
to the graph of Fig.~\ref{fig:NRQEDgraphs}a reads
\begin{eqnarray}
{\cal{M}}_e&=& \sum\hspace{-5.5mm}\int\limits_{n,\mathbf{P}} \,
{ \me{0}{ A^{(2\gamma)}}{n,\mathbf{P}} 
i \me{n,\mathbf{P}}{{ie  \over m} \left\{ \mathbf{p}_1 \cdot \mathbf{A}(\mathbf{x}_1)-\mathbf{p}_2\cdot\mathbf{A}(\mathbf{x}_2) \right\}}
{\text{o-Ps}}
\over E_o-E_{n,\mathbf{P}}-\omega}\,.
\label{1.4}
\end{eqnarray}
The sum above extends over all discrete and continuum states $| n,\mathbf{P} \rangle$ of the spectrum
of the unperturbed Hamiltonian $H_0$, that can be written as the direct product of a plane wave with
the c.m. momentum $\mathbf{P}$ and a wavefunction describing the relative motion.
In configuration space, $\langle \mathbf{x}_1\mathbf{x}_2 | n,\mathbf{P} \rangle = 
e^{i\mathbf{P}\cdot\mathbf{X}}\,\psi_n(\mathbf{x})$,
where $\psi_n(\mathbf{x})$ is an eigenstate of $H_C$ with eigenvalue $E_n$,
and we have omitted the spin wavefunction.
The energy of this intermediate (virtual) $e^+e^-$ state is $E_{n,\mathbf{P}}=E_n+\mathbf{P}^2/4m$, 
and $E_0=-m\alpha^2 /4$ is the o-Ps ground state energy.

%%%%%%%%%%%%%%%%%%%%%%%%%%%%%%%%%%
%%%%%%%%%%%%%%%%%%%%%%%%%%%%%%%%%%

The first matrix element in the r.h.s of Eq.~(\ref{1.4}) gives the short-distance part
of the annihilation. The quantity $A^{(2\gamma)}$ is the $e^+e^-\to 2\gamma$ amplitude in QED
calculated as an expansion in the momenta of the leptons:
\begin{eqnarray}
A^{(2\gamma)}
&\equiv& \chi^\dagger_{-\mathbf{p}} \left(\mathrm{W}_0 + \mathbf{W}_1 \cdot \mathbf{p} + \mathbf{W}_2 \cdot \mathbf{P}\right)\phi_{\mathbf{p}}
+ {\cal O}(\mathbf{p}^2)\,,
\label{1.5}
\end{eqnarray}
where $\phi_{\mathbf{p}},\,\chi^\dagger_{-\mathbf{p}}$ are the Pauli spinor fields that annihilate the electron and positron with relative momentum with respect the c.m. $\mathbf{p}$ and $-\mathbf{p}$,
respectively. Only the first order in the momentum expansion will be needed in this work. Expressions
for $\mathrm{W}_0,\,\mathbf{W}_1$ can be found in Ref.~\cite{our}, while 
$\mathbf{W}_2=(\mathbf{W}_1+(\bm{\epsilon}_1\cdot \bm{\epsilon}_2)\,\bm{\sigma})/2$,
with $\bm{\epsilon}_1,\, \bm{\epsilon}_2$ the polarizations of the outgoing hard photons.
The $\mathbf{W}_2\cdot \mathbf{P}$ term in Eq.~(\ref{1.5}) is an operator which 
depends on the total momentum of the $e^+e^-$ pair.
In principle, we need to consider the latter because in the o-Ps
decay the intermediate $e^+e^-$ pair can recoil after the soft photon is radiated. Inserting the
operators in Eq.~(\ref{1.5}) into the short-distance amplitude one gets:
\begin{eqnarray}
\me{0}{ A^{(2\gamma)}}{n,\mathbf{P}} = \mbox{Tr} \left[ \big( \mathrm{W}_0\,\psi_n(0) 
- i\,\mathbf{W}_1 \cdot \left\{ \mathbf{\nabla}\,\psi_n(\mathbf{y})\right\}_{\mathbf{y}=0}
+ \mathbf{W}_2\cdot \mathbf{P}\,\psi_n(0) \big) 
\big( \phi\,\chi^{\dagger} \big)
 \right] \,.
 \label{1.6}
\end{eqnarray}
The trace above is taken over spinor indices, and $\phi\,\chi^{\dagger}$ 
is the spin wavefunction of the state $|n,\mathbf{P}\rangle$,
which can be in a spin-0 state ($\phi\,\chi^{\dagger}\to {\mathbf 1}/\sqrt{2}$), or in a spin-1 state with polarization vector $\bm{\varepsilon}$ ($\phi\,\chi^{\dagger}\to \bm{\varepsilon}^*\cdot \bmsigma /\sqrt{2}$).

The matrix element for the radiation of the soft photon with three-momentum $\mathbf{k}$ and
polarization vector $\bm{\epsilon}$ (we pick a gauge where photon polarizations are purely transverse,
$\epsilon_i\cdot k_i=0$, $\epsilon_i^0=0$) can be calculated easily:
\begin{eqnarray}
\me{n,\mathbf{P}}{ \mathbf{p}_1 \cdot \mathbf{A}(\mathbf{x}_1)
-\mathbf{p}_2\cdot\mathbf{A}(\mathbf{x}_2) }{\text{o-Ps}}
&=& -i\,(2\pi)^3 \delta^{(3)}(\mathbf{P}+\mathbf{k})\,(\bm{\varepsilon} \cdot \bm{\varepsilon}_0^*)\nn\\
&&\times
\int d^3\mathbf{x}\,\psi_n^*(\mathbf{x})
\left( e^{-i\mathbf{k}\cdot\mathbf{x}/2}+e^{i\mathbf{k}\cdot\mathbf{x}/2} \right)
\bm{\epsilon}\cdot\nabla\psi_0(\mathbf{x})\,,
\label{1.7}
\end{eqnarray}
with the initial o-Ps state $\langle \mathbf{x}_1\mathbf{x}_2|\text{o-Ps}\rangle \equiv\psi_0(\mathbf{x})$,
once we set $\mathbf{X}=0$.
$\psi_0(\mathbf{x})$ is the ground state wave function in position space 
$$
\psi_0(\mathbf{x})={1 \over \sqrt{\pi a^3} }\,e^{-x/a}\,,
$$
and $\bm{\varepsilon}_0$ is the spin-1 polarization vector. The photon wavefunctions $e^{\mp i\bmk\cdot\bmx/2}$
arise from the emission of the photon from the electron or from the positron line, respectively.
The $\mathbf{p}\cdot\mathbf{A}$ interaction can change the orbital angular momentum but not the
spin of the $e^+e^-$ pair, which thus remains in a spin-1 state. The delta function in Eq.~(\ref{1.7}) sets the recoil
momenta of the intermediate $e^+e^-$ system to $-\mathbf k$, so that $E_{n,\mathbf{P}}=E_n+\omega^2/4m$ is its
energy~\footnote{Let us recall 
that in TOPT three-momentum is conserved at the vertices and virtual states are always on-shell.}.
Taking into account the sum over polarizations,
$$
\sum \bm{\varepsilon}^*_i\bm{\varepsilon}_j = \delta_{ij}\,,
$$
we can write 
\begin{eqnarray}
{\cal{M}}_e &=& - \frac{e}{m} \,
\langle  \chi^\dagger\, \mathrm{W}_1^i  \, \phi\rangle_{\bm{\varepsilon}_0}
\int d^3\mathbf{x}\,
\left\{ \sum_n\frac{\mathbf{\nabla}_{\mathbf{y}}^i\psi_n(\mathbf{y})\,\psi_n^*(\mathbf{x})}
{E_{n,\mathbf{k}}+\omega-E_0}\right\}_{\mathbf{y}=0}
\!\!\!\!\!\!
\left( e^{-i\mathbf{k}\cdot\mathbf{x}/2}+e^{i\mathbf{k}\cdot\mathbf{x}/2} \right)
\bm{\epsilon}\cdot\nabla\psi_0(\mathbf{x})\nn\\[2mm]
&&\!\!\!\!\!\!\!\!\!\!\!\!\!\!\!\!\!\!+ \,\frac{i \, e}{m}\,
\langle  \chi^\dagger\,  \mathbf{W}_2 \cdot  \, \mathbf{k} \, \phi\rangle_{\bm{\varepsilon}_0}
\int d^3\mathbf{x}
\left\{ \sum_n\frac{\psi_n(0) \, \psi_n^*(\mathbf{x})}{E_{n,\mathbf{k}}+\omega-E_0} \right\}
\left( e^{-i\mathbf{k}\cdot\mathbf{x}/2}+e^{i\mathbf{k}\cdot\mathbf{x}/2} \right)
\bm{\epsilon}\cdot\nabla\psi_0(\mathbf{x})\,.
\label{1.8}
\end{eqnarray}
The term $\mathrm{W}_0$ in Eq.~(\ref{1.6}) has disappeared from the electric amplitude because
it only contributes to the $2\gamma$ annihilation of a spin-singlet (parapositronium: p-Ps) state. Likewise, the second line in Eq.~(\ref{1.8}) vanishes 
because the vector integration can only be proportional to the photon momentum $\mathbf{k}$ 
and $\mathbf{k}\cdot\bm{\epsilon}=0$. Therefore we are left only with the term proportional to
$\mathbf{W}_1$, where the term between brackets can be written in terms of the Coulomb Green's function at
energy $-\kappa^2/m=E_0-\omega-\omega^2/4m$,
\begin{eqnarray}
\left\{ \sum_n\frac{\mathbf{\nabla}_{\mathbf{y}}^i\psi_n(\mathbf{y})\,\psi_n^*(\mathbf{x})}
{E_{n,\mathbf{k}}+\omega-E_0} \right\}_{\mathbf{y}=0} 
\!\!\!\!\!= \, \left\{ \mathbf{\nabla}_{\mathbf{y}}^i G(\mathbf{y},\mathbf{x};\kappa) \right\}_{\mathbf{y}=0}
= \, 3\,\mathrm{x}^i\,G_1(0,r;\kappa)
\,.\label{1.9}
\end{eqnarray}
The derivative acting on the Green's function picks out the $\ell=1$ component of its partial wave decomposition
in $\mathrm{y}=0$,
\begin{eqnarray}
G\left( \mathbf{x}, \mathbf{y}, k \right)  &=& \sum_{\ell=0}^\infty \left(2 \ell +1 \right) \left( x y \right)^\ell
P_\ell \left( \mathbf{x \cdot y} / x y \right) G_\ell \left( x, y ,k \right)\,.
\label{1.10}
\end{eqnarray}
(See Appendix~C of Ref.~\cite{our} for explicit expressions of the partial waves $G_\ell$). Therefore only $P$-wave
states contribute in the sum over virtual fluctuations in Eq.~(\ref{1.4}).
Using $\nabla\psi_0(\mathbf{x})=-\psi_0(\mathbf{x})\,\mathrm{x}^i/a\,r$, we can rewrite the electric amplitude as
\begin{eqnarray}
{\cal{M}} &=& - e\, \psi_0(0)\, 
\langle  \chi^\dagger\, \mathrm{W}_1^i  \, \phi\rangle_{\bm{\varepsilon}_0}
%\mbox{Tr} \left[ \, \mathrm{W}_1^i \, { \bmsigma \cdot \bm{\varepsilon}^{*}_0 \over \sqrt{2}} \, \right]
\, \epsilon^j\, D^{ij}(\mathbf{k})\,,
\label{1.11}
\end{eqnarray}
with
\begin{eqnarray}
D^{ij}(\mathbf{k})=-\frac{3}{ma}\int d^3\mathbf{x}\,\frac{\mathrm{x}^i \mathrm{x}^j}{r}\,
\left( e^{-i\mathbf{k}\cdot\mathbf{x}/2}+e^{i\mathbf{k}\cdot\mathbf{x}/2} \right)
G_1(0,r;\kappa) \, e^{-r/a}
\,.
\label{1.12}
\end{eqnarray}
The general structure of $D^{ij}(\mathbf{k})$ reads
\begin{eqnarray}
D^{ij}(\mathbf{k})
= {\cal{D}}(\omega)\,\delta^{ij} + \widetilde{\cal{D}}(\omega)\frac{\mathrm{k}^i\mathrm{k}^j}{\omega^2}\,.
\label{1.13}
\end{eqnarray}
The term proportional to $\widetilde{\cal{D}}(\omega)$ vanishes when contracted with the photon polarization $\bm{\epsilon}$.
The coefficient ${\cal{D}}(\omega)$ can be projected out by contracting with the tensor $\delta^{ij}-\mathrm{k}^i\mathrm{k}^j/\omega^2$,
and after performing the angular integration, we obtain:
\begin{eqnarray}
{\cal{D}}(\omega)=-\frac{8\pi}{ma} \int dr \,r^3\, G_1(0,r;\kappa) \, e^{-r/a} \, 
\left[ \, {24\over (\omega r)^3}\sin \left( \frac{\omega r}{2} \right) 
-{ 12\over (\omega r)^2} \cos\left( \frac{\omega r}{2} \right) \right] 
\,.
\label{1.14}
\end{eqnarray}
Note that all recoil effects up to this order are properly accounted for by the 
$\omega^2/4m$ term in the parameter $\kappa$ that enters the Green's function, and are naturally
suppressed with respect leading terms in the non-relativistic regime.

In addition, gauge invariance requires that the momentum operators in the
short-distance amplitude
$A^{(2\gamma)}$ are replaced by the covariant derivatives
$\mathbf{p}_i \mp e\mathbf{A}(\mathbf{x}_i)$~\cite{our}. This generates a purely local term that
has to be added to the electric amplitude calculated with the interaction Hamiltonian of Eq.~(\ref{1.1}).
The corresponding graph is depicted in Fig.~\ref{fig:NRQEDgraphs}b. This  contribution reads
\begin{eqnarray}
\me{0}{ A^{(2\gamma)} } {\text{o-Ps}} = 
- e \, \psi_0(0) \, 
\langle  \chi^\dagger\,  \mathbf{W}_1 \cdot \bm{\epsilon} \, \phi\rangle_{\bm{\varepsilon}_0}
%\mbox{Tr} \left[ \big( \mathbf{W}_1 \cdot \bm{\epsilon} \big)\, { \bmsigma \cdot \bm{\varepsilon}^{*}_0 \over \sqrt{2}} \right] 
\,,
 \label{1.15}
\end{eqnarray}
and the total electric amplitude is then equal to
\begin{eqnarray}
{\cal{M}}_e &=& - e\, \psi_0(0)\,
\langle \chi^\dagger \,\mathbf{W}_1 \cdot \bm{\epsilon} \, \phi\rangle_{\bm{\varepsilon}_0}
\Big( 1+{\cal{D}}(\omega)  \Big)
\,.
\label{1.16}
\end{eqnarray}

The expression for ${\cal{D}}(\omega)$ in Eq.~(\ref{1.13}) allows us to identify the typical
distances that control the falloff of the integrand. The $\ell=1$ Green's function,
\begin{eqnarray}
G_1(0,r;\kappa)&=&{m \kappa^3 \over 3 \pi}e^{-\kappa r }
\,\Gamma(2-\nu)\,U(2-\nu,4,2\kappa r)\quad,\quad\nu=\frac{m\alpha }{2\kappa}=\frac{1}{a\kappa} \,,
\label{1.17}
\end{eqnarray}
contains the exponential factor $e^{-\kappa r}$.
On the other hand, we have the factor $e^{-r/a}$ from the wavefunction of the initial state.
Therefore, distances in the integrand for the electric amplitude are constrained by the exponential factor
\begin{eqnarray}
\exp \left\{-\Big(\kappa+1/a\Big) r \right\}\,,
\label{1.18}
\end{eqnarray}
and the qualitative behaviour of ${\cal{D}}(\omega)$ depends on the scaling of the soft photon energy:\\

\noindent
$\bullet \; \omega \sim m\alpha^2 \;(\kappa\sim m\alpha)$:
both exponential factors are of the same size, constraining the integration variable
to $r\sim a$. The multipole expansion is applicable here as the product $\omega r\sim \alpha$, and
it is realized in the full result~(\ref{1.14}) by expanding the trigonometric functions between
brackets:
\begin{eqnarray}
 {24\over (\omega r)^3}\sin \left( \frac{\omega r}{2} \right) 
-{ 12\over (\omega r)^2} \cos\left( \frac{\omega r}{2} \right)
\,=\,1-\frac{(\omega r)^2}{40}+{\cal O}(\omega^4r^4)\,,
\label{1.19}
\end{eqnarray}
and thus
\begin{eqnarray}
{\cal{D}}(\omega)&=&d_e(\omega)+d_e^{(1)}(\omega)+\ldots\nn\\[2mm]
&=&d_e(\omega)
\left( 1+{\cal{O}} \left( \omega^2/a^2 \right) \right)
\,,
\label{1.20}
\end{eqnarray}
where the leading term is given by
\begin{eqnarray}
d_e(\omega)=-\frac{8\pi}{ma} \int dr \, r^3 \, G_1(0,r;\kappa) \, e^{-r/a} 
\,.
\label{1.21}
\end{eqnarray}
The quantity 
$$
1+d_e(\omega)\equiv a_e(\omega)
$$ 
gives the electric amplitude in the dipole limit, which was calculated in Ref.~\cite{our} (Eq.~(60) therein) using an interaction Hamiltonian written in a gauge invariant form in terms of the electric field~\footnote{One has to approximate $\kappa \simeq \sqrt{\omega-mE_0}$ to make contact with the result in Refs.~\cite{our,voloshin}, since
recoil corrections were not considered there.}. A formula
for $a_e(\omega)$ in terms of a hypergeometric function was given in Ref.~\cite{voloshin}, and can also be found in the
Appendix~\ref{sec:appen1}. 
In this region the 
Coulomb interaction among the $e^+e^-$ pair after soft photon radiation has a strength given by $\nu\sim {\cal{O}}(1)$, and is resummed to all orders through the $\ell=1$ Coulomb Green's function. Note that the 
usual QED perturbative expansion does not accommodate binding effects and cannot be applied to obtain the 
photon spectrum in this region.   
For smaller photon energies, $\omega\ll m\alpha^2$ the parameter $\nu\to 1$ and $a_e(\omega)\simeq 2\omega/m\alpha^2$, in agreement with Low's theorem prediction for this decay~\cite{our}.\\

\noindent
$\bullet \; \omega \sim m\alpha \;(\kappa\sim \sqrt{m\omega})$: since now $\kappa \gg 1/a$, the falloff of 
${\cal{D}}(\omega)$ is ruled by the exponential factor from the Green's function, so the characteristic
distance is $r\sim \kappa^{-1} \sim 1/m\alpha^{1/2}$. The argument of the trigonometric functions
is constrained to $\omega r \sim \sqrt{\omega /m} \sim \alpha^{1/2}$, which allows to multipole expand the full result also
for radiated photons of this energy. The photon wavefunctions can be expanded out from the beginning
and the leading term in such expansion is again given by $d_e(\omega)$ above, although
higher terms yield now corrections proportional to $(\omega/m)^n$:
\begin{eqnarray}
{\cal{D}}(\omega)&=&d_e(\omega)+d_e^{(1)}(\omega)+\ldots\nn\\[2mm]
&=&d_e(\omega)
\left( 1+{\cal{O}} \left( \omega/m \right) \right)
\,.
\label{1.22}
\end{eqnarray}
A new feature arises in the whole region $m\alpha^2\ll \omega \ll m$ because the parameter
$\nu\simeq \frac{\alpha}{2}\sqrt{\frac{m}{\omega}}$ is small and allows to further expand the 
Coulomb Green's function inside $d_e(\omega)$:
\begin{eqnarray}
G_1(0,r;\kappa) \; \xrightarrow{\nu\to 0} \;
{m \over 12 \pi}
\,\frac{1+\kappa r}{r^3}\,e^{-\kappa r }\equiv \, G^{f}_1(0,r;\kappa)
 \,,
\label{1.23}
\end{eqnarray}
{\it i.e.} the leading term of the expansion in $\nu$ of the $P$-wave Coulomb Green's function
is equal to the $P$-wave projection of the free Green's function. The latter shows that we 
can treat the Coulomb interaction as a perturbation in this region, and higher order terms in the
$\nu$-expansion of $G_1$ correspond to insertions of the Coulomb potential in the diagrammatic picture. 
Using the result (\ref{1.23}) and expanding out the o-Ps wave function $e^{-r/a}$, one obtains the leading term in $\nu$
of $d_e(\omega)$,
\begin{eqnarray}
d_e(\omega) &=&
-\frac{8\pi}{ma} \int dr \, r^3\, G^{f}_1(0,r;\kappa)  + \ldots \,=\,
-\frac{4}{3}\nu+{\cal{O}} (\nu^2) \nn\\
&=&-\frac{2\alpha}{3}\sqrt{\frac{m}{\omega}}\left(1-\frac{\omega}{8m}-\frac{m\alpha^2}{8\omega}+\dots \right)
+{\cal{O}} (\nu^2)
\,,
\label{1.24}
\end{eqnarray}
which agrees with the $\nu$-expansion of the exact result of $d_e(\omega)$, which can be found in 
Eq.~(\ref{A.9}) of the Appendix~\ref{sec:appen1}. A $\omega/m$ correction
to the lowest order result is generated automatically in the ${\cal O}(\nu)$ term
of Eq.~(\ref{1.24}) from the recoil correction which is included in 
$\kappa$. Higher order terms in the $\nu$ expansion correspond to the neglected
terms in the expansion of $G_1$ and the o-Ps wavefunction.
The next term in the expansion 
of the photon wavefunctions, $d_e^{(1)}(\omega)$ in Eq.~(\ref{1.20}), also produces a leading
term proportional to $\nu(\omega/m)\simeq \alpha/2\sqrt{\omega/m}$, which is 
obtained by replacing $G_1 \to G_1^{f}$ and $e^{-r/a}\to 1$:
\begin{eqnarray}
d_e^{(1)}(\omega)&=&
\frac{8\pi}{ma} \int dr \, r^3\, G_1(0,r;\kappa) \, e^{-r/a} \,\frac{(\omega r)^2}{40} \nn\\[2mm]
&=&\frac{\omega^2\pi}{5ma} \int dr \, r^5\, G^{f}_1(0,r;\kappa)  + \ldots =\frac{2}{15}\nu\left(\frac{\omega}{\kappa}\right)^2 +{\cal{O}} \Big(\nu^2\frac{\omega^2}{\kappa^2}\Big)
\,.
\label{1.25}
\end{eqnarray}
The relation
\begin{eqnarray}
\left(\frac{\omega}{\kappa}\right)^2 =\frac{\omega}{m}\,\frac{1-\nu^2}{1+\frac{\omega}{4m}}= \frac{\omega}{m}+\dots
\label{1.252}
\end{eqnarray}
can be used in Eq.~(\ref{1.25}) to write $d_e^{(1)}(\omega)$ as an expansion in $\omega/m$.
The exact result for $d_e^{(1)}(\omega)$ can be found in the Appendix~\ref{sec:appen1}.
For completeness we also give there the exact result of the full electric  amplitudes ${\cal{D}}(\omega)$.

Corrections
of ${\cal{O}}(\omega/m)$ can also arise considering higher powers in the momentum expansion of the
annihilation amplitude $A^{(2\gamma)}$. The higher order terms in $\mathbf{p}$ will pick intermediate $e^+e^-$ pairs
with higher angular momentum $L$ when they are transformed into derivatives acting on the Green's function. The scaling
of this part of the electric amplitude is of order 
$$
\langle 0| \, \mathbf{p}\left( \frac{\mathbf{p}}{m} \right)^{L-1}G \, | \mathbf{x}\rangle
\sim 
\left( \frac{\kappa}{m} \right)^{L-1} \langle 0| \, \mathbf{p} \, G \, | \mathbf{x}\rangle\,,
$$
{\it i.e.} shows a suppression $\sim (\omega/m)^{\frac{L-1}{2}}$ with respect the $L=1$ case, since 
$\mathbf{p}\sim \kappa \sim \sqrt{\omega m}$.
On the other hand, the radiative transition to the intermediate $e^+e^-$ $L$-wave state,
${}^3S_1 \to\gamma + {}^3L_J$,
will select a $L$-tensor 
in the multipole expansion of the photon wavefunctions, that will yield a 
$(\omega r)^{L-1}\sim (\omega /m)^{\frac{L-1}{2}}$ extra factor with respect the lowest order transition amplitude ($L=1$). 
Combining both factors we see that higher order terms in the expansion of $A^{(2\gamma)}$ for the electric amplitude are thus down by an overall factor $(\omega/m)^{L-1}$. In order to compute these corrections it is convenient to write the expanded amplitude of $A^{(2\gamma)}$ in terms of the spherical harmonics built from the vector $\mathbf{p}$, so they can project the required $L$-wave component of the Green's function directly. A representation of the spherical harmonics in terms of the Cartesian coordinates of
the vector is specially suited for this task, and can be found in Ref.~\cite{cartesian}.

Let us now turn to the magnetic contribution. We will not detail all the steps as we did for the electric amplitude, but
just quote the result:
\begin{eqnarray}
{\cal{M}}_m &=& \sum\hspace{-5.5mm}\int\limits_{n,\mathbf{P}} \,
{ \me{0}{ A^{(2\gamma)}}{n,\mathbf{P}} 
i \me{n,\mathbf{P}}{{i \mu} \left\{ \bm{\sigma}_{\phi}\cdot\mathbf{B}(\mathbf{x}_1)+
\bm{\sigma}_{\chi}\cdot\mathbf{B}(\mathbf{x}_2) \right\}}
{\text{o-Ps}}
\over E_o-E_{n,\mathbf{P}}-\omega}
\nn\\
&=& -i \frac{e}{m} \, \psi_0(0)\,\mathrm{W}_0\,\bm{\delta}\cdot 
\langle \chi^\dagger \bm{\sigma} \, \phi\rangle_{\bm{\varepsilon}_0}
\, {\cal{A}}(\omega)
\,,
\label{1.26}
\end{eqnarray}
with
\begin{eqnarray}
{\cal{A}}(\omega)= 8\pi \int dr \,r\, G_0(0,r;\kappa) \, e^{-r/a} \, 
\sin \left( \frac{\omega r}{2} \right) 
\,,
\label{1.27}
\end{eqnarray}
and $\bm{\delta}=({\mathbf{k}}\times\bm{\epsilon})/\omega$..
Photon emission through a 
$\bmsigma\cdot \mathbf{B}$ term
can induce a transition from a triplet to a singlet spin configuration.
The constant term in the annihilation amplitude $A^{(2\gamma)}$
allows only for intermediate $e^+e^-$ states with quantum numbers ${}^1S_0$, which are contained in
the $\ell=0$ component of the Green's function. As in the case of the electric amplitude, the falloff of the
Green's function fixes the characteristic distance for the multipole expansion to $r\sim\kappa^{-1}$, so that
we can always expand in $\omega r$ for $\omega\ll m$. 
Under the dipole approximation, the photon wavefunctions are dropped 
and the radiative matrix element reduces to the
scalar product between the ground state and the rest of Coulomb
wavefunctions, which are mutually orthogonal. 
In that limit, only the ${}^1S_0$ (p-Ps) ground state contributes~\cite{our}, whose energy differs 
from that of the ground state o-Ps only if the hyperfine splitting is considered. The latter is only 
relevant for the study of the o-Ps decay spectrum at very low photon energies, $\omega\sim m\alpha^4$~\cite{our},
and can be neglected here. The multipole expansion of the magnetic amplitude ${\cal{A}}(\omega)$ then reads:
\begin{eqnarray}
{\cal{A}}(\omega)&= &4\pi \omega \int dr \,r^2\, G_0(0,r;\kappa) \, e^{-r/a} \,\left(1-\frac{(\omega r)^2}{24}+\ldots\right)
\nn \\
&=& a_m(\omega) + a_m^{(1)}(\omega) +\ldots
\nn \\[2mm]
&=& a_m(\omega) \left( 1+{\cal{O}}({\omega/m}) \right)
\,,
\label{1.28}
\end{eqnarray}
where the magnetic amplitude in the dipole approximation,
\begin{eqnarray}
a_m(\omega) = \frac{1}{1+\omega/4m} 
\,,
\label{1.29}
\end{eqnarray}
is trivially equal to one if recoil corrections are neglected. If we take into account higher terms
in the expansion of the photon wavefunction, transitions between the ground state and $S$-wave radial
excitations are possible, and the Coulomb interaction becomes relevant. For photons with energies
in the range $m\alpha^2 \ll \omega \ll m$ the Coulomb interaction can be treated perturbatively and
the first ${\cal O}(\alpha)$ correction arises in the subleading term in the multipole expansion,
$a_m^{(1)}(\omega)$, and is 
proportional to $\nu(\omega/m)\simeq \alpha/2\sqrt{\omega/m}$ (see Eq.~(\ref{A.15})).
The exact result for $a_m^{(1)}(\omega)$ and for ${\cal{A}}(\omega)$ are given in the Appendix~\ref{sec:appen1}.

According to Eqs.~(\ref{1.20},\ref{1.28}), the NRQED calculation of the o-Ps decay amplitude at leading order in the multipole expansion reads:
\begin{eqnarray}
{\cal{M}}= {\cal{M}}_e+{\cal{M}}_m= - e\, \psi_0(0)\,
\left(
\langle \chi^\dagger \,\mathbf{W}_1 \cdot \bm{\epsilon} \, \phi\rangle_{\bm{\varepsilon}_0}\, a_e(\omega)
+i  \, \frac{\mathrm{W}_0}{m}\,\bm{\delta}\cdot 
\langle \chi^\dagger \bm{\sigma} \, \phi\rangle_{\bm{\varepsilon}_0}
\, a_m(\omega)
\right)
\,.
\label{1.30}
\end{eqnarray}
From a knowledge of the amplitude, we can derive an expression for the o-Ps differential photon spectrum
at low energies ($\omega\ll m$)
written in terms of the electric and magnetic amplitudes~\cite{our}:
\begin{eqnarray}
{ {\rm d}\Gamma \over  {\rm d}x }  &=&{m \alpha^6 \over 9 \pi} x \left[ \abs{a_m}^2 +  \frac 7 3 \abs{a_e}^2\right] 
\quad,\quad x\equiv \omega/m\,,
\label{1.31}
\end{eqnarray}
Let us now focus on the region $\omega\sim m\alpha$.
It is clear that since in this region Coulomb effects can be accounted for perturbatively,
one can apply the conventional QED loop expansion. The leading (LO) and next-to-leading (NLO) terms in
$\alpha$ of the $\text{o-Ps}\to 3\gamma$ NRQED
amplitude when $x\sim \alpha$ are obtained by taking 
\begin{eqnarray}
a_e=1- { 2 \over 3}{\alpha \over \sqrt{x}}+{\cal O}(\alpha) \quad  ,\quad a_m=1+{\cal O}(\alpha) \quad 
\,,
\label{1.32}
\end{eqnarray}
and the NRQED spectrum in this region, up to NLO, thus reads:
\begin{eqnarray}
{ {\rm d}\Gamma \over  {\rm d}x }  &=&{2 m \alpha^6 \over 27 \pi} 
\left[ 5x -\frac{14}{3}\,\alpha\sqrt{x} +  {\cal{O}}(\alpha^{2}) \right]
\,.
\label{1.33}
\end{eqnarray}
The LO term of the NRQED amplitude~(\ref{1.30}) matches the $\omega \to 0$ limit of the tree-level QED calculation first done by Ore and Powell~\cite{orepowell}, as it was shown in Ref.~\cite{our}. The same is of course also true
for the LO photon spectrum.
The comparison for the $\alpha\sqrt{x}$ correction requires the
one-loop result of the QED series. 
In Sec.~\ref{sec:QED} we will derive the soft photon
 limit of the 1-loop QED spectrum obtained from an analytic expression of the phase-space distribution
given in Ref.~\cite{adkins}, and check that the leading term when $\omega\to 0$ agrees with the 
$\alpha\sqrt{x}$ correction in the NRQED spectrum shown above~\footnote{A comparison between the effective theory ${\cal O}(\alpha)$ spectrum 
and a numerical calculation of the one-loop QED result~\cite{adkinsnum} has been done by Voloshin~\cite{voloshin}, showing 
a good agreement.}. In Sec.~\ref{sec:threshold} we will compute the leading term of the 1-loop QED amplitude
for soft photon energy by means of the threshold expansion, and show that it arises from a loop-momentum region
$\mathbf{p} \sim \sqrt{\omega m}$, and that it indeed agrees with the NRQED result.
Concerning higher orders in the $\omega/m$
expansion for the ${\cal O}(\alpha^0)$ and ${\cal O}(\alpha)$ terms discussed above,
any possible mismatch between the QED and the NRQED 
amplitudes would require that we introduce a short-distance $e^+e^-\to 3\gamma$ local
contribution in the NRQED Hamiltonian~\footnote{Let us remind that the NRQED Hamiltonian is constructed 
to reproduce the QED amplitude neglecting binding effects.}.

\section{One-loop QED spectrum for $\bm{m\alpha^2\ll \omega\ll m}$}
\label{sec:QED}

An analytic evaluation of the 1-loop $\text{o-Ps}\to 3\gamma$ decay amplitude has not become available until recently~\cite{adkins}. In the later work, Adkins has provided a compact form for the one-loop phase-space distribution,
which allows to obtain the one-loop correction to the energy spectrum upon integration:
\begin{eqnarray}
{ {\rm d}\Gamma_1 \over  {\rm d}x_1 }  &=&{m \alpha^7 \over 36 \pi^2} 
\int_{1-x_1}^1 dx_2 \frac{1}{x_1 x_2 x_3} \left\{ F(x_1,x_2)+ \text{permutations} \right\} \,,
\label{2.1}
\end{eqnarray}
where $x_i=\omega_i/m\,(i=1,2,3)$, with $\omega_i$ the energy of the photons, which satisfy that $x_1+x_2+x_3=2$.
Analytic expressions for the function $F(x_1,x_3)$ can be found in the Appendix of Ref.~\cite{adkins}.
They involve a number of intermediate functions of the variables $x_i$, including dilogarithm functions.
Therefore an analytic integration of the formula~(\ref{2.1}) may be difficult to obtain. However,
for an evaluation of the low-energy spectrum 
one does not need to perform the full integration before taking the limit $x_1\to 0$.
A Taylor series in $x_1$ can be obtained by first taking the small photon energy limit of
the integrand in Eq.~(\ref{2.1}) up
to the desired accuracy, and afterwards doing the (trivial) integration. Since also the
variables $1-x_2$ and $1-x_3$ are small when $x_1\to 0$ , some care is needed in order
to perform this limit.

The two essential variables that parameterize the $1\to 3$ phase-space have been choosen in Eq.~(\ref{2.1})
to be $x_1$ and $x_2$, where $x_1$ is the variable of the differential spectrum. Therefore $x_1\ll x_2\lsim 1$.
Note that in the $x_2$ integration the limits imply that $1-x_2< x_1$, so we can write $1-x_2\equiv  x_1 t$, with $t\in [0,1]$, and change the integration variable to $t$:
\begin{eqnarray}
{ {\rm d}\Gamma_1 \over  {\rm d}x_1 }  &=&{m \alpha^7 \over 36 \pi^2} 
\int_0^1 dt \,\frac{\widetilde{F}(x_1,t)}{(1-x_1 t)(1-x_1(1- t))} 
\,,
\label{2.2}
\end{eqnarray}
where
\begin{eqnarray}
\widetilde{F}(x_1,t)=F(x_1,1-x_1 t)+ \text{permutations}
\,.
\label{2.3}
\end{eqnarray}
The integrand in Eq.~(\ref{2.2}) is symmetric under the change $t\leftrightarrow (1-t)$, which is indeed equivalent to 
changing $x_2 \leftrightarrow x_3$. With the redefinition above we can  Taylor expand  the integrand in the variable $x_1$ while keeping the variable $t$ fixed, and the resulting terms turn out to be well behaved in the $x_1\to 0$ limit.
Up to the second
non-trivial order it yields a relatively simple phase-space distribution:
\begin{eqnarray}
\frac{\widetilde{F}(x_1,t)}{(1-x_1 t)(1-x_1(1- t))} &=& -\frac{16\pi}{3}\left[ 3-4t(1-t) \right] \sqrt{x_1} \nn\\
&& +
\left[ \, 2t(1-t)\left( 3\pi^2-48\log 2+40 \right)+\pi^2+16\log 2 -60  \, \right]x_1 \nn\\[2mm]
&& + \, {\cal{O}}(x_1^{3/2})
\,,
\label{2.4}
\end{eqnarray}
that can also be written back in terms of $x_2$ if needed.
Accordingly, the 1-loop QED correction to the low-energy spectrum reads (writing $x_1\to x$)
\begin{eqnarray}
{ {\rm d}\Gamma_1 \over  {\rm d}x }  &=&{2 m \alpha^6 \over 27 \pi} 
\left[ -\frac{14}{3}\alpha\sqrt{x} + \frac{\alpha}{\pi}\left( \frac{3\pi^2}{4}-\frac{35}{2} \right) x+ {\cal{O}}(\alpha x^{3/2}) \right]
\,.
\label{2.5}
\end{eqnarray}
We see that the $\alpha\sqrt{x}$ term above agrees with the corresponding one in the NRQED formula, Eq.~(\ref{1.33}).
The $\alpha x$ correction  also matches the corresponding term in the EFT computation, 
which was included in the formula~(\ref{1.33})
by Voloshin~\cite{voloshin} considering the ${\cal{O}}(\alpha)$ correction to the electron gyromagnetic
ratio and the ${\cal{O}}(\alpha)$ corrections to the annihilation amplitude $e^+e^-\to 2\gamma$. 
The latter correction can be accounted for in a straightforward way if
the NRQED spectrum is written in terms of local four-fermion operators which effectively
integrate out all possible states of hard photons~\cite{our}. In this description,
the coefficients multiplying the squares of the electric and magnetic amplitudes in the 
photon spectrum~(\ref{1.31}), are identified with the imaginary part of the matching coefficients 
of $S$- and $P$-wave four-fermion annihilation operators (see Eqs.~(21,74) of Ref.~\cite{our}).
These coefficients can be found up to ${\cal{O}}(\alpha^3)$ in Ref.~\cite{bbl}, and yield
a part of the $\alpha x$ correction shown above. The remaining part, due to the radiative correction
to the electron gyromagnetic ratio, is accounted for in the NRQED Lagrangian naturally as the
1-loop matching coefficient of the $\bm{\sigma}\cdot\mathbf{B}$ interaction.

To conclude, we wish to point out that higher order corrections in the $x$-expansion of the 1-loop low-energy
photon spectrum can be computed straightforwardly from the analytic phase-space distribution given by Adkins through
the expansion procedure outlined here.

\section{Threshold expansion of the QED 1-loop diagrams}
\label{sec:threshold}

%%%%%%%%%%%%%%%%%%%%%%%%%%%%%%%%%%
\begin{figure}
\begin{center}
\includegraphics[width=12cm]{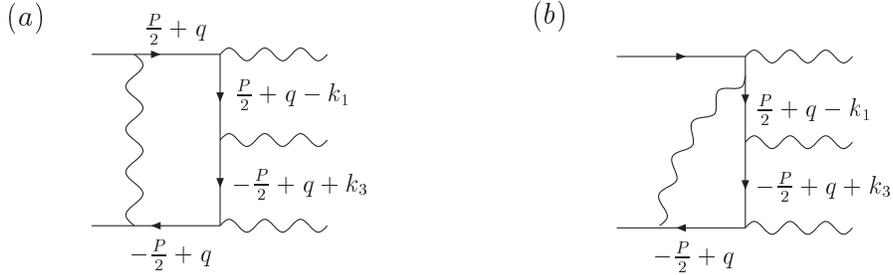}
\end{center}
\caption{Ladder and double-vertex graphs contributing to the o-Ps decay amplitude
to order $\alpha$. Momenta flow in the same direction as the arrows. \label{fig:ladder}}
\end{figure} 
%%%%%%%%%%%%%%%%%%%%%%%%%%%%%%%%%%
The threshold expansion technique is a prescription developed by
Beneke and Smirnov~\cite{beneke} for the asymptotic expansion of loop integrals
involving heavy massive particles close to threshold ({\it i.e.} that are moving at
small velocities). It provides the velocity expansion of an amplitude up to a
certain order as a set of simpler integrals than those present in the original Feynman diagram.
The method relies on the so-called strategy of regions, which replaces the integration
over the whole space of loop momenta to the integration only over some specific 
regions. The first and crucial step is to identify the relevant momentum regions 
in the loop integrals. These in principle follow from the
pole structure of the Feynman propagators, that is characterized by the 
relevant scales that appear in the problem. For amplitudes involving
a non-relativistic pair, three scales enter the dynamics: the
mass of the particle, $m$, the relative 3-momentum 
$\sim m v$ and the non-relativistic energy $\sim m v^2$.
Accordingly, four different loop momentum regions were
identified to give non-zero contributions in Ref.~\cite{beneke}:
\begin{eqnarray}
\text{hard}\ :&&  q_0 \, \sim \, \mathbf{q} \, \sim \, m\,,
\nonumber\\[3mm]
\text{soft}\ :&&  q_0 \, \sim  \, \mathbf{q} \, \sim \, m v\,,
\nonumber\\[3mm]
\text{potential}\ :&& q_0 \, \sim \, \mathbf{q}^2/m \, \sim mv^2\,,
\nonumber\\[3mm]
\text{ultrasoft}\ :&& q_0 \, \sim \, \mathbf{q}  \, \sim \,  mv^2\,.
\label{3.1}
\end{eqnarray}    

In a second step the original integration region must be decomposed into a sum of
integrals, one for every region above, and a Taylor 
expansion in the parameters that are small in each regime performed. 
Every integral, containing just one scale, thus contributes 
only to a single power in the velocity 
expansion, which can be determined easily before integration.
The procedure requires the use of dimensional regularization,
even if the original integration is finite~\footnote{Other analytic regularization procedures can be employed,
since the important point is that scaleless integrals are put to zero.}.
Following these heuristic rules, the authors of 
Ref.~\cite{beneke} reproduced the exact $v$ expansion of some one-loop
and two-loop examples. Although a formal proof of the validity of
the asymptotic expansion close to threshold has not been given, the perfect
agreement in the examples supports their use in general one-loop diagrams.

We now wish to apply the threshold expansion to the 1-loop $\text{o-Ps}\to 3\gamma$ 
amplitude when one of the photons has an energy $\omega \sim m\alpha$, where a conventional
perturbative QED calculation is still valid. The momentum of the emitted soft photon, $k\sim m\alpha$,
introduces a further scale in the amplitude and, as we will show below, gives rise to a new region that contributes to the exact result. It corresponds to a loop momentum
\begin{eqnarray}
  q_0 \, \sim \, \mathbf{q}^2/m \, \sim \, \omega\,,
\label{3.2}
\end{eqnarray}
which arises from the heavy particle propagators and indeed resembles the potential region
but with a different scaling for the 3-momentum $\mathbf{q}\sim \sqrt{m\omega}$. It thus corresponds
to a small momentum region ($q\ll m$) which lies 
in between the hard relativistic region and the small momentum regions shown in Eq.~(\ref{3.1}), and 
we shall refer to it in the following as `soft-radiation' region. 
The physical origin of the new region
has been discussed in the NRQED description of the o-Ps decay in Sec.~\ref{sec:NRQED}.

The 1-loop graphs contributing to the $\text{o-Ps}\to 3\gamma$ amplitude can be classified 
into one of the following categories~\cite{adkinsannals,adkins}: self-energy, vertex corrections, double-vertex, ladder and annihilation contributions. The leading term in the $\omega/m$ expansion is given by the ladder and double-vertex graphs, shown in 
Figs.~\ref{fig:ladder}a and~\ref{fig:ladder}b, respectively. The momenta flowing through the fermion propagators have been written in the figure. The Positronium atom four-momentum is $P=k_1+k_2+k_3$, with components $P=(M,0)$ in the o-Ps c.m. frame, and we can take $M\simeq 2m$ safely
for the discussion that follows. Note that the assignment of momenta in the figure implies that the electron and positron inside the bound state have incoming momenta $P/2 = (m,0)$. The evaluation
of the 1-loop diagrams for the o-Ps decay with $e^+e^-$ strictly at threshold is a valid procedure to 
obtain the matching coefficients of the NRQED annihilation operators that give
the o-Ps total decay rate, as it has been shown in Ref.~\cite{adkinsannals}. Setting the velocity
to zero automatically kills off the contributions of the small momentum regions shown in Eq.~(\ref{3.1}), 
so that only hard and soft-radiation momentum regions will survive in the QED 1-loop o-Ps amplitude~\footnote{Loosely
speaking, in the approach of Ref.~\cite{adkinsannals} 
the non-relativistic dynamics is properly taken into account by the NRQED part of the calculation.}.
In order to show that the threshold expansion technique for diagrams with a soft photon emission does not only apply
for this particular kinematic configuration, we will consider in the Appendix~\ref{sec:appen2} a more general case 
with $P^2\ne 4m^2$, 
where several small momentum regions exist at the same time.

Let us consider the ladder graph shown in 
Fig.~\ref{fig:ladder}a, where the soft photon is the one attached to the incoming electron. Therefore $k_1\sim m\alpha$,
while $k_2,\,k_3$ are hard photons scaling as $m$. 
Two integrals turn out to be relevant in the $\omega_1 \to 0$ limit of the ladder amplitude. The first one is the 5-point
scalar integral:
\begin{eqnarray}
I_0  =  \int \!\!
{ [d^Dq] \over  q^2 \, [(q+P/2)^2-m^2] \, [(q-P/2)^2-m^2] \,  [(q+P/2-k_1)^2-m^2] \, [(q-P/2+k_3)^2-m^2] }\,.
\nn\\
\!\!\!\!\!\! \label{3.3}
\end{eqnarray}
The integration measure is defined as $[d^Dq]\equiv d^Dq/i(2\pi)^{D}$, in $D=4-2\epsilon$ dimensions, and the
standard $+i\epsilon$ prescriptions are implicitly understood in the propagators. When the loop momentum is hard,
we can expand the fermion propagator in $k_1$. In this way, only the scale $m$ enters into the propagators in the hard
region. Taking into account that $[d^Dq]\sim m^4$, the leading term in the contribution from the hard region
is estimated to be $I_0^{(\text{h})}\sim m^{-6}$.
However, for $\omega_1\to 0$ the dominant contribution to the integral above comes from a
loop-momentum region where $q\ll m$. Since the momentum that flows through the fermion propagator 
connecting the two hard photons is always hard, when $q\ll m$ we can expand
\begin{eqnarray}
{1 \over [(q-P/2+k_3)^2-m^2]}= {1 \over -P\cdot k_3}
-\frac{q\cdot P-2q\cdot k_3}{(P\cdot k_3)^2}+\dots  
\label{3.4}
\end{eqnarray}
so that the propagator effectively shrinks to a local annihilation vertex. Likewise,
in the remaining massive 
propagators we can drop the $q_0^2$ terms with respect $q\cdot P=2m q_0$. The leading contribution of $I_0$
in the small momentum region is thus
\begin{eqnarray}
I_0^{(\text{small})}  = -\frac{1}{2m\omega_3} \int 
{ [d^Dq] \over  (q_0^2-\mathbf{q}^2) \, (2m q_0-\mathbf{q}^2) \, (-2m q_0-\mathbf{q}^2) \,
(2m q_0-2m\omega_1-\mathbf{q}^2) \ }\,.
\label{3.5}
\end{eqnarray}
We have also dropped the term $q\cdot k_1 \ll 2mq_0$. Closing the integration contour in the upper complex
$q_0$-plane, we pick the contribution from the pole of the massive propagator at $q_0=-\mathbf{q}^2/2m+i\epsilon$. 
The contribution of the residue of this pole to $I_0^{(\text{small})}$ reads 
\begin{eqnarray}
I_0^{(\text{s-r})}  = -\frac{i}{16 m^2\omega_3} \int [d^n\bmq] \,
{1 \over  (\mathbf{q}^2)^2 \, (\mathbf{q}^2+m\omega_1)  \,  } = \frac{1}{64\pi m^2\omega_3} (m\omega_1)^{-\frac{3}{2}}
+{\cal{O}}(\epsilon)\,.
\label{3.6}
\end{eqnarray}
with $n=D-1$.
In deriving Eq.~(\ref{3.6}) we have also expanded the photon propagator in $q_0^2$, since $q_0^2\ll \mathbf{q}^2$
once we pick the residue from the fermion pole. The integral in Eq.~(\ref{3.6}) is finite and it is dominated by loop momentum $\mathbf{q}\sim \sqrt{m\omega}$, which corresponds to the scaling of the soft-radiation loop momentum region introduced above
\footnote{The integral in Eq.(45) with $\omega_1$ replaced by $\Lambda_{\text{QCD}}$ was also found in a non-perturbative enviroment~\cite{brambilla}.}. 
As in the hard region, the contribution from the soft-radiation region to the final result can be readily
estimated before making any integration by first expanding the integrand according to the 
hierarchy $q_0\sim\mathbf{q}^2/m\sim m\omega_1$. For the leading term the propagators yield
a factor $\sim m^{-2}(m\omega_1)^{-4}$, while $[d^Dq]\sim  m^{-1}(\mathbf{q}^2)^{5/2}\sim m^{-1}(m\omega_1)^{5/2}$, so that
we obtain an overall scaling $I_0^{(\text{s-r})} \sim m^{-3}(m\omega_1)^{-3/2}$, as found in Eq.~(\ref{3.6}). 
The exact result of the integral $I_0$ can be easily derived from the 4- and 5-point functions computed in 
Ref.~\cite{adkins}. We have checked that the leading term in the $\omega_1\to 0$ limit of the
exact expression for $I_0$ agrees with $I_0^{(\text{s-r})}$ above.
There is also a contribution to $I_0^{(\text{small})}$ from the pole of the massless propagator at
$q_0=-|\mathbf{q}|+i\epsilon$. After picking up the corresponding residue the remaining
integration is dominated by loop-momenta $|\mathbf{q}|\sim \omega_1$. The 
characteristic loop-momentum in this region is thus the same as for the soft region in Eq.~(\ref{3.1}),
if we consider $v\sim \alpha$. The contribution
from this pole has an overall scaling $[d^Dq]/(m q_0)^5\sim m^{-5}\omega_1^{-1}$, {\it i.e.} larger
than the hard region one, although suppressed by $\sqrt{m\omega_1}$ with respect the soft-radiation region.
We do not give the
result because we will not keep terms of the similar order that arise from other diagrams.

The second integral that we shall need for the leading term in the $\omega_1\to 0$ limit of the 
ladder amplitude is the 4-point tensor integral
\begin{eqnarray}
I_0^{ij}  =  \int[d^Dq] \,
{  q^iq^j\over  q^2 \, [(q+P/2)^2-m^2] \, [(q-P/2)^2-m^2] \,  [(q+P/2-k_1)^2-m^2] }%\, [(q-P/2+k_3)^2-m^2] }
\,,
\label{3.7}
\end{eqnarray}
with the indices $i,j=1,2,3$. We have already expanded out the hard fermion propagator in order to write 
down $I_0^{ij}$, since the leading contribution will come up from the $q\ll m$ regime. 
As we shall see when we put back the numerators of the amplitudes, 
part of the $q^iq^j$ term arises when a $q^j$ is brought to the numerator of the integrand from
the subleading term in the $q\ll m$ expansion of the hard propagator shown in Eq.~(\ref{3.4}).
Strictly, one should do the tensor reduction of the integral and afterwards apply the threshold expansion to the scalar integrals that appear. However, since the tensor structure in the integral in Eq.~(\ref{3.7}) only involves the spatial indices, we shall take the path of expanding the integrand according to the different loop-momenta
directly without prior reduction to scalar integrals. In the small loop-momentum region, the propagators in 
$I_0^{ij(\text{small})}$ are identical to those in $I_0^{(\text{small})}$, and thus they share the same pole structure. The expansion of the propagators in the soft-radiation region leads to:
\begin{eqnarray}
I_0^{ij\,(\text{s-r})}  = \frac{i}{8 m} \int [d^n\bmq]
{q^iq^j \over  (\mathbf{q}^2)^2 \, (\mathbf{q}^2+m\omega_1)  \,  } = 
\frac{1}{96\pi m}\frac{\delta^{ij}}{\sqrt{m\omega_1}} 
+{\cal{O}}(\epsilon)
\,.
\label{3.8}
\end{eqnarray}
After dropping the $\mathbf{k}_1$-dependence in the fermion propagator, the reduction of the tensor integral is trivial. 
The contribution from the gluon pole in $I_0^{ij}$ can be shown to scale as $\sim  \omega_1/m^3$,
and it is thus suppressed with respect the soft-radiation contribution.

For the double vertex diagram of Fig.~\ref{fig:ladder}b the leading term in the $\omega_1\to 0$ limit comes from
the integral
\begin{eqnarray}
I_1  =  \int \!\!
{ [d^Dq] \over  q^2 \,  [(q-P/2)^2-m^2] \,  [(q+P/2-k_1)^2-m^2] }%\, [(q-P/2+k_3)^2-m^2] }
\,,
\label{3.9}
\end{eqnarray}
after expanding the hard region propagator as done in Eq.~(\ref{3.4}). 
The expansion of $I_1$ in the small loop-momentum gives an analogous expression to (\ref{3.5}) but with one fermion
propagator less. The soft-radiation region arises from the fermion pole at $q_0=-\mathbf{q}^2/2m+i\epsilon$, and yields
\begin{eqnarray}
I_1^{(\text{s-r})}  = -\frac{i}{4 m} \int [d^n\bmq]
{1 \over  \mathbf{q}^2 \, (\mathbf{q}^2+m\omega_1)  \,  } = -\frac{1}{16\pi m} (m\omega_1)^{-\frac{1}{2}}
+{\cal{O}}(\epsilon)
\,.
\label{3.10}
\end{eqnarray}
On the other hand, the contribution from the gluon pole in the small loop-momentum scales as $m^{-2}$, so it is again
suppressed with respect the soft-radiation region.

\begin{figure}
\begin{center}
\includegraphics[width=12cm]{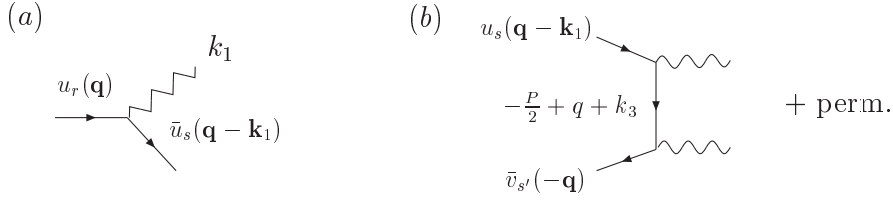}
\end{center}
\caption{The soft photon vertex (${\cal M}_1$) and the $2\gamma$ annihilation (${\cal M}_{23}$) parts in the ladder graph.\label{fig:soft+2gamma}}
\end{figure}

Let us now include the numerators of the propagators and vertex factors in order to calculate the full amplitude. 
We can use the identity
\begin{eqnarray}
\frac{\xslash{p}+m}{p^2-m^2+i\epsilon}=
 \frac{\sum_s u_s(\mathbf{p}) \bar{u}_s(\mathbf{p})}{p^0-E_{\mathbf{p}}+i\epsilon}
  +  \frac{\sum_{s} v_{s}(-\mathbf{p}) \bar{v}_{s}(-\mathbf{p})}{p^0+E_{\mathbf{p}}-i\epsilon}  
\label{3.11}
\end{eqnarray}
with $E_\bmp=\sqrt{m^2+\bmp^2}\simeq m+\bmp^2/(2m)$, to pick the numerators associated to the non-relativistic fermion poles 
that are relevant when the loop-momentum is small, and afterwards perform the non-relativistic expansion of the 
spinors in $\mathbf{p}/m$. This will automatically lead us to the usual non-relativistic Feynman rules for the vertices
and propagators. The identity (\ref{3.11}) effectively separates particle and antiparticle contributions, as it is conventional
in the time-ordered perturbation theory. The spinors in Eq.~(\ref{3.11}) have non-relativistic normalization, more precisely:
\begin{eqnarray}
u_s(\bmp) = \sqrt{\frac{E_\bmp+m}{2E_\bmp}}
\left( \begin{array}{c} \phi_s \\ \frac{\bm{\sigma}\cdot\bmp}{m+E_{\mathbf{p}}}
\,\phi_s \end{array} \right)\;\;\;,
\;\;\;
v_{s}(\bmp) = \sqrt{\frac{E_\bmp+m}{2E_\bmp}}\left( \begin{array}{c} 
\frac{\bm{\sigma}\cdot\bmp}{m+E_\bmp}\,\chi_{s} \\
\chi_{s}  \end{array} \right)
\,,
\label{3.12}
\end{eqnarray}
where $s$ is the spin label. Let us illustrate how the non-relativistic expansion is performed for the ladder diagram.
The structure of the fermion propagators in the small loop-momentum region has been shown in Eq.~(\ref{3.5}). Take, for example,
the fermion propagator at the left of the soft photon vertex, with momentum $q+P/2$. The expansion of this propagator
for $q\ll m$ yields $2mq^0 -\bmq^2$. The latter agrees (up to a normalization factor) with the first denominator that is obtained
from the identity~(\ref{3.11}), which reads $q^0+P^0/2-E_\bmq \simeq q^0-\bmq^2/2m$. Therefore we will associate the electron
spinors $\sum_s u_s(\bmq) \bar{u}_s(\bmq)$ to this denominator in order to derive the leading term in the small momentum region.
In the same way, the expanded fermion propagators after soft photon emission and after the $k_3$ vertex correspond to electron
and positron contributions, respectively, in the splitting of Eq.~(\ref{3.11}). For the massive line connecting 
the $k_2$ to the $k_3$ vertices, we need to keep the full covariant propagator because the momentum flowing through it is always hard. Taking into account the particle and antiparticle spinor wavefunctions from the adjacent propagators, the non-relativistic 
reduction of this part of the ladder diagram is indeed equivalent to the $e^+e^-\to 2\gamma$ annihilation amplitude of
an incoming electron with 3-momentum $\bmq-\mathbf{k}_1$ and an incoming positron with 3-momenta $-\bmq$ (see Fig.~\ref{fig:soft+2gamma}b), 
\begin{eqnarray}
{\cal M}_{23} = -i\,e^2\,\bar{v}_{s^\prime}(-\bmq)\,\xslash{\epsilon_3}\;\,
\frac{1}{\xslash{q}-\xslash{P}\,/2+\xslash{k}_3-m}\;
\xslash{\epsilon_2}\;\,u_s(\bmq-\mathbf{k}_1)+\{ 2\leftrightarrow 3\}\,,
\label{3.13}
\end{eqnarray}
where we have also included the contribution from the graph with the hard photons permuted. 
Recall that we found already the non-relativistic expansion of the $e^+e^-\to 2\gamma$ amplitude 
in Sec.~\ref{sec:NRQED}. Indeed, the expansion of Eq.~(\ref{3.13}) for the hierarchy
$m\gg \mathbf{q} \gg \mathbf{k}_1$, characteristic of the soft-radiation region, yields the same leading
and subleading terms already shown in Eq.~(\ref{1.5}), so that
\begin{eqnarray}
{\cal M}_{23}
= \chi^\dagger_{s^\prime} \left(\mathrm{W}_0 + \mathbf{W}_1 \cdot \mathbf{q} \right)\phi_s
+ {\cal O}(\alpha)\,.
\label{3.14}
\end{eqnarray}
The size of the terms in the expansion of ${\cal M}_{23}$ above should be estimated considering 
the soft-radiation region scaling, $\mathbf{q}\sim \sqrt{m\omega_1}\sim m\alpha^{1/2}$. The reason why
we only need to keep the two first terms in ${\cal M}_{23}$ shall become clear when we put together
all the pieces in the amplitudes.
%%%%%%%%%%%%%%%%%%%%%%%%%%%%%%%%%%
%%%%%%%%%%%%%%%%%%%%%%%%%%%%%%%%%%
Similarly, the non-relativistic expansion of the soft-photon vertex is also done easily once we pick the 
particle contribution from the propagators (see Fig.~\ref{fig:soft+2gamma}a):
\begin{eqnarray}
{\cal M}_1 = -i\,e\,\bar{u}_s(\bmq-\mathbf{k}_1)\,\xslash{\epsilon_1}\;u_r(\bmq)
= i\frac{e}{m}\,\bmq\cdot\bm{\epsilon}_1\,\phi^\dagger_s\phi_r+\frac{e}{2m} \, (\bmk_1\times\bm{\epsilon}_1) \,\phi^\dagger_s \bmsigma\phi_r
+{\cal O}(\alpha^{3/2})\,.
\label{3.15}
\end{eqnarray}
We can check that the non-relativistic reduction of the amplitude ${\cal M}_1$ agrees with the Feynman rule
that is derived for the $e^-\to e^- + \gamma$ vertex from the non-relativistic interaction Hamiltonian, Eq.~(\ref{1.1}).
Finally, the photon exchange between the initial $e^+e^-$ reduces to a Coulomb-like potential:
\begin{eqnarray}
{\cal M}_c = 
\frac{i e^2}{q^2}\,\bar{u}_r(\bmq) \gamma^\mu u(\mathbf{0})\,\bar{v}(\mathbf{0}) \gamma_\mu v_{s^\prime}(-\bmq)
= -\frac{i e^2}{\bmq^2} \, \phi^\dagger_r \phi \,\chi^\dagger\chi_{s^\prime}
+{\cal O}(\alpha^0)\,,
\label{3.16}
\end{eqnarray}
in the region $q_0\sim \bmq^2/m \ll m$, and with $\phi,\,\chi$ the Pauli spinors of the initial $e^+e^-$. We can
now put together the different pieces, ${\cal M}_c$, ${\cal M}_1$ and ${\cal M}_{23}$, including the denominators
and taking into account that $\sum_s\phi_s\phi^\dagger_s=\sum_s\chi_s\chi^\dagger_s=\bmone$. The leading term of the
threshold expansion of the ladder amplitude in Fig.~\ref{fig:ladder}a (plus the graph with the photons $k_2$ and $k_3$
interchanged) is thus equal to
\begin{eqnarray}
{\cal M}^{\text{ladder}} &=& 
\frac{ e^3}{m}\, \int [d^n q]
{ \langle \chi^\dagger \, \mathbf{W}_1 \cdot\bmq \, \phi \rangle\, ( \bmq\cdot \bm{\epsilon}) 
-i\mathrm{W}_0 \,\bm{\delta}_1\cdot 
\langle \chi^\dagger \bm{\sigma} \, \phi\rangle\over 
(\mathbf{q}^2) \, (q_0-\mathbf{q}^2/2m) \, (q_0+\mathbf{q}^2/2m) \,
(q_0-\omega_1-\mathbf{q}^2/2m)  } 
 \nn\\[2mm]
& = & e\,\langle \chi^\dagger \, \mathbf{W}_1 \cdot\bm{\epsilon} \, \phi \rangle \,
\frac{\alpha}{3}\sqrt{\frac{m}{\omega_1}}
+i e \, \frac{\mathrm{W}_0}{m}  \,\bm{\delta}_1\cdot 
\langle \chi^\dagger \bm{\sigma} \, \phi\rangle
\,
\frac{\alpha}{2}\sqrt{\frac{m}{\omega_1}}\,.
\label{3.17}
\end{eqnarray}
The integrals appearing in Eq.~(\ref{3.17}) are proportional to the soft-radiation
contributions of the integrals $I_0$ and $I_0^{ij}$, computed in Eqs.~(\ref{3.6}) and (\ref{3.8}),
respectively. We can proceed in a similar way to obtain the leading term in the threshold
expansion of the double-vertex diagram, Fig.~\ref{fig:ladder}b:
\begin{eqnarray}
{\cal M}^{\text{dou.ver.}} &=& 
\frac{ i e^3}{2m}\, \mathrm{W}_0 \,\bm{\delta}_1\cdot 
\langle \chi^\dagger \bm{\sigma} \, \phi\rangle\int
{[d^n q] \over 
(\mathbf{q}^2) \,  (q_0+\mathbf{q}^2/2m) \,
(q_0-\omega_1-\mathbf{q}^2/2m)  } 
 \nn\\[2mm]
& = & -i e \, \frac{\mathrm{W}_0}{m} \,\bm{\delta}_1\cdot 
\langle \chi^\dagger \bm{\sigma} \, \phi\rangle
\, \frac{\alpha}{2}\sqrt{\frac{m}{\omega_1}}\,,
\label{3.18}
\end{eqnarray}
where we have required the contribution from the soft-radiation region of $I_1$, Eq.~(\ref{3.10}).
For the double-vertex amplitude there is no contribution from the $\bmq\cdot \bm{\epsilon}_1$
term in the soft-photon vertex because the incoming electron is static. The $\mathrm{W}_0$
terms from the ladder and double-vertex graphs cancel each other and we are only left with 
a $\alpha\sqrt{m/\omega_1}$ proportional to $\mathbf{W}_1$. It is straightforward to verify that 
the contribution from the ladder
graphs where the soft photon vertex is attached in the lower part of the diagram ({\it i.e.} to the incoming
positron line) gives the same $\mathbf{W}_1$ term. Therefore, the leading term in $\omega_1\to 0$ limit
of the 1-loop $e^+e^-\to 3\gamma$ QED amplitude reads
\begin{eqnarray}
{\cal M}^{\text{1-loop}} 
& = & e\,\langle \chi^\dagger \, \mathbf{W}_1 \cdot\bm{\epsilon} \, \phi \rangle \,
\frac{2\alpha}{3}\sqrt{\frac{m}{\omega_1}}
\,,
\label{3.19}
\end{eqnarray}
which agrees with the leading ${\cal O}(\alpha)$ term in the NRQED computation of this amplitude, Eqs.~(\ref{1.30}) 
and~(\ref{1.32}), once we include the o-Ps wavefunction factor. 

A comment on the rest of 1-loop diagrams that had been not considered is appropriate. First
one should note that all diagrams where the soft photon $k_1$ is placed among the two hard ones scale
as ${\cal O}(1)$ because the momenta flowing through the fermion propagators that connect two photon vertices 
are always hard. With respect the annihilation contribution, the scaling can be inferred from that of the ladder 
graph if we replace the $t$-channel photon exchanged between the initial $e^+e^-$ by a $s$-channel 
photon exchange. The replacement carries a $\bmq^2/m^2$ suppression in the non-relativistic expansion, 
which is of order $\omega_1/m$ for the soft-radiation contribution. 
Regarding the self-energy and vertex correction topologies, they correspond
to typical relativistic corrections whose effects are not expected to be enhanced when we explore 
the non-relativistic regime. Indeed it can be easily checked from an quick analysis of the propagators in the small
loop-momentum region that for the integrals in the vertex and self-energy corrections, 
we can always avoid the fermion poles so that only
the pole in the photon propagator at $q^0\sim \bmq$ contributes. In this way, the small loop-momentum 
expansion of the 2- and 3-point scalar integrals give results which are down by $(\omega_1/m)^2$ and $\omega_1/m$, respectively, with respect the hard region contributions to the same integrations.

%%%%%%%%%%%%%%%%%%%%%%%%%%%%%%%%%%
\begin{figure}
\begin{center}
\includegraphics[width=12cm]{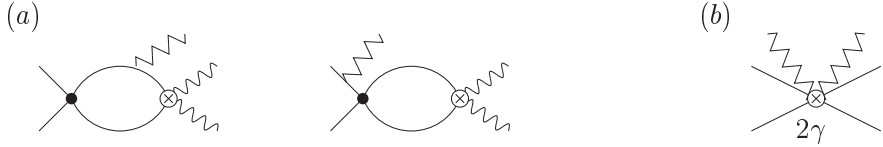}
\end{center}
\caption{$(a)$ NRQED diagrams that contribute to the ${\cal O}(\alpha\sqrt{m/\omega_1})$ term of the $e^+e^-\to 3\gamma$
decay amplitude. The zigzag lines represent soft photons; $(b)$ local vertex that describes real soft photon radiation from
the heavy fermion-antifermion pair in vNRQED.
The $2\gamma$ label denotes that the vertex arises from the integration of two hard photons. \label{fig:EFT1loop}}
\end{figure} 
%%%%%%%%%%%%%%%%%%%%%%%%%%%%%%%%%%

It is interesting to note that the expressions~(\ref{3.17},\ref{3.18}) correspond to the 
amplitudes that would be written down from the diagrams displayed in Fig.~\ref{fig:EFT1loop}a using the usual
NRQED Feynman rules with free fermions as asymptotic 
states~\footnote{Recall that we used bound state perturbation theory in coordinate space
to compute the NRQED amplitude in Sec.~\ref{sec:NRQED}.
In the $\omega\sim m\alpha$ region 
the leading ${\cal O}(\alpha)$ correction in the latter computation
could be obtained by replacing
the intermediate Coulomb Green's function by the free one and neglecting the o-Ps wavefunction dependence
(see Eq.~(\ref{1.24})). This is 
effectively equivalent to consider the Coulomb interaction as a perturbation, so that in this regime the NRQED computation
can be done using as unperturbed states either free propagating fermions (as in Fig.~\ref{fig:EFT1loop}a)
or $e^+e^-$ Coulomb eigenstates (as in Sec.~\ref{sec:NRQED}).},
and after we drop the $\mathbf{k}_1$ term in the fermion 
propagator:
\begin{eqnarray}
{1 \over q^0-\omega_1-(\bmq-\mathbf{k}_1)^2/2m}\simeq {1 \over q^0-\omega_1-\bmq^2/2m}\,.
\label{3.20}
\end{eqnarray}
Keeping
the full NRQED propagator and expanding afterwards leads of course to the same leading order term, but 
unnecessarily complicates the calculation.
In the usual NRQED counting for bound state systems, the heavy fermion and antifermion
inside the bound state have energies $E\sim m\alpha^2$ and relative 3-momentum $\bmq \sim m\alpha$, so 
they are classified as potential degrees of freedom. As we explained in Sec.~\ref{sec:NRQED}, 
the radiation of a soft photon from the $e^+e^-$ allows for
higher virtualities in the relative 3-momentum, of order $\bmq\sim\sqrt{m\omega_1}$.
Therefore the heavy particles propagating in the loops of Fig.~\ref{fig:EFT1loop}a do not obey 
the potential scaling (otherwise the massive propagator (\ref{3.20}) for a potential fermion will reduce
to just $1/\omega_1$ and we will miss the contribution from the soft-radiation region).
The fact that we have to further expand the NRQED propagators in the soft-radiation regime
is an unwanted feature because it jeopardizes power-counting.
However, we can take advantage from the fact that soft radiation takes place at distances $\sim 1/\sqrt{m\omega}$, 
smaller that the typical size
of the Coulomb bound state, to integrate out the contributions stemming from the
soft-radiation momentum region. From the point of view of the potential fermions,
soft radiation is a short-distance process, just like the $2\gamma$-annihilation, and can be
described by a local vertex. 
In the vNRQED/vNRQCD~\cite{vNRQCD} formalism (with the external sources for the hard photons integrated out)
the decay spectrum in the soft-energy can be calculated from the imaginary part of the matrix element
of a 6-field
operator as the one shown in Fig.~\ref{fig:EFT1loop}b. The matching coefficient of such operator
can be obtained 
by comparing with the QED result in the $\omega_1\sim m\alpha$ region: at tree-level it is
just a constant$\times \alpha^3$ (the limit $x\to0$ of the Ore-Powell decay spectrum~\cite{orepowell}, divided by the $x$ factor
from the phase-space), while at ${\cal O}(\alpha)$ it is proportional to $\alpha^4\sqrt{m/\omega}$.
Note that this picture is similar to the NRQCD factorization approach~\cite{bbl} used to describe the production and annihilation 
of heavy quarkonium, that we briefly discuss in the next section.

\section{Relevance for heavy quarkonium}
\label{sec:quarkonium}

The threshold expansion technique applied above for the o-Ps$\to 3 \gamma$ decay, can
also be useful to calculate the direct photon spectrum of quarkonium radiative
decays in the soft-energy region. Heavy quarkonium decay rates can be understood
within the NRQCD factorization approach~\cite{bbl}, which allows to separate the short distance 
physics related with the heavy quark-antiquark annihilation process from the 
long distance bound state dynamics. An operator product expansion can be written
down for the direct photon spectrum in, for example, $\Upsilon$ decays:
\begin{eqnarray}
{{\rm d}\Gamma \over {\rm d}x} = \sum_n C_n(x) \langle \Upsilon|{\cal O}_n | \Upsilon\rangle\,,
\label{4.1}
\end{eqnarray}
where the sum above extends over all $Q\bar{Q}[n]$ configurations that can be found inside
the quarkonium, and the $C_n(x)$
are short distance Wilson coefficients that can be determined from the annihilation
cross section of the on-shell $Q\bar{Q}[n]$ pair as a perturbative series in $\alpha_s(m_b)$.
The ${\cal O}_n$ in the long-distance matrix element are NRQCD operators which are
organized in powers of the relative velocity of the heavy quarks.
At leading order
only the color-singlet operator ${\cal O}_1({}^3S_1)$, that creates and annihilates
a quark-antiquark pair in a color-singlet ${}^3S_1$ configuration, contributes. The
nonperturbative NRQCD matrix element is related to the $\Upsilon$ wave function at the origin
\begin{eqnarray}
\langle \Upsilon|{\cal O}_1({}^3S_1) | \Upsilon\rangle
= \langle \Upsilon | \phi_{\bmp}^\dagger \bmsigma^i \chi_{-\bmp} \, \chi^{\dagger}_{-\bmp^\prime} \bmsigma^i \phi_{\bmp^\prime} 
 | \Upsilon\rangle=2N_c\,|\psi(0)|^2
\,,
\label{4.2}
\end{eqnarray}
where now $\phi_{\bmp},\chi_{-\bmp}\,$ refer to heavy quark and antiquark fields respectively.
The matching coefficient at leading order in $\alpha_s$ comes from the tree-level $Q\bar{Q}\to gg \gamma$
annihilation~\cite{upsilon},
\begin{eqnarray}
C_1(x) &=& {32\over 27}{ \alpha\alpha_s^2 Q^2_b \over m_b^2 } 
\left[\frac{2 - x}{x} 
+ \frac{\left( 1 - x\right) \,x}{{\left( 2 - x \right) }^2} 
- \frac{2\,{\left( 1 - x \right) }^2\,\log (1 - x)}{{\left( 2 - x \right)}^3}
 + \frac{2\,\left( 1 - x\right) \,\log (1 - x)}{x^2} \right]\,,
\nn\\
\label{4.3}
\end{eqnarray}
which has identical $x$-dependence to that of the Ore-Powell spectrum for o-Ps~\cite{orepowell}.
Indeed the factorization approach applies trivially to QED bound states, where everything
is calculable perturbatively and the bound state wavefunction reduces to the Coulomb one.
The leading order coefficient for Ps decays, $\widetilde{C}_1(x)$, is obtained 
by the trivial replacements $\alpha_s^2 Q_b^2/m_b^2 \to \alpha^2/m^2$ and $32/27\to 16/9$ in
Eq.~(\ref{4.3}). In the soft-energy region, it reads
\begin{eqnarray}
\widetilde{C}_1(x) &=&  {40\over 27}{ \alpha^3 \over m^2 }\, x+{\cal O}(x^2) 
\,,
\label{4.4}
\end{eqnarray}
The ${\cal O}(\alpha)$ correction to the o-Ps$\to 3\gamma$
amplitude for soft photon energies computed in the previous
sections can also be included, so that 
\begin{eqnarray}
\widetilde{C}_1^{\text{NLO}}(x) &=& 
 {8\over 27}{ \alpha^3 \over m^2 }\left[
5x-{14\over 3}\,\alpha\sqrt{x}  +{\cal O}(\alpha x^0) \right]
\,,
\label{4.5}
\end{eqnarray}
that we can plug in Eq.~(\ref{4.1}) to obtain the o-Ps photon spectrum to NLO
in the soft-energy region (we also need to replace $\psi\to \psi_c$ and $N_C\to 1$ in Eq.~(\ref{4.2})). 
A similar result to that of Eq.~(\ref{4.5}) could be obtained for the 
${\cal O}(\alpha\alpha_s^3)$ short-distance coefficients $C_n(x)$
in heavy quarkonium decays by doing the threshold expansion
of the corresponding 1-loop QCD diagrams. The latter could provide a
determination of the ${\cal O}(\alpha_s)$ corrections to the quarkonium photon spectrum in the soft-energy range of photon 
energies. To our knowledge, $\alpha_s$ corrections to the coefficient 
$C_1(x)$ in Eq.~(\ref{4.3}) are only known numerically~\cite{kraemer}.
We should be aware though that the fragmentation contributions to the photon spectrum in quarkonium decays
({\it i.e.} those in which the photon is emitted from the decay products) become
important in the low-$x$ region~\cite{catani,maltoni}, namely for $x\lsim 0.3$ in $\Upsilon$ decays,
and have to be properly taken into account for a computation of the full spectrum. At lower energies,
$x\lsim \alpha_s^2$, the emission of the photon can produce transitions to virtual bound states
and does not longer belong to the short-distance part of the decay~\cite{our}. The standard NRQCD factorization also 
breaks down at large values of the photon energy~\cite{rothstein}, where one needs to consider also collinear
degrees of freedom~\cite{upsiloncoll}.

\section{Conclusions}
\label{sec:conclusion}

The region of photon energies $\omega\sim m\alpha$ in the radiative decay of a heavy
fermion-antifermion pair has been analyzed in this work through the study of the three-photon
decay of ground state orthopositronium. The soft-energy region 
effectively separates the regime where binding effects become essential,
from the hard-energy region, where the details of the bound state dynamics are irrelevant.
The NRQED framework is able to yield
the correct o-Ps spectrum from energies $\omega < m\alpha^2$, where it properly accounts for
bound state effects, up to energies $\omega\sim m\alpha$, where the binding can be neglected and
the calculation is simplified by replacing the Coulomb Green's function with the free one.
We have found agreement in the comparison of 
the o-Ps spectrum in the soft-energy region as
computed with NRQED and with conventional perturbation theory in QED.
For the NRQED computation, the leading order approximation in the multipole expansion of the photon 
field yields already the dominant contribution in the whole energy range $\omega\ll m$, 
as it had been argued by Voloshin~\cite{voloshin}.
We have explicitly calculated the contribution from higher order multipoles to the o-Ps decay amplitude
in NRQED, and discussed the size of relativistic corrections coming from other sources in the soft-energy
region. 

The 1-loop QED result of the o-Ps spectrum in the soft-energy region
has been obtained by two different methods. First by expanding
the analytic expression for the phase-space distribution obtained by Adkins~\cite{adkins}, and second
by means of the threshold expansion technique. With the latter method it has been shown
that the dominant contribution arises from a new loop-momentum region $q^0\sim \bmq^2/m \sim m\alpha$.
This momentum region arises naturally when the
heavy fermion-antifermion system decays emitting a soft photon
and has to be added to the rest of momentum regions which are known to be relevant for
the analysis of heavy particle-antiparticle loop diagrams. The identification of 
momentum regions that have not been previously considered is mandatory for the success of the asymptotic 
expansion method in a wider range of kinematic situations. It is also of conceptual importance for
the construction of 
EFT's for these systems with well-defined
power-counting rules and able to reproduce the correct low-energy behaviour.

\begin{acknowledgments} 
I have benefited from numerous discussions on the subject of this work with A.~Hoang and
A.~Manohar. I also thank them for their comments on the manuscript. 
I am particularly grateful to G.~Adkins for sharing the details of his calculation and for very useful correspondence.
Part of this work was done at the Instituto de F\'\i sica Corpuscular de Valencia (IFIC).
This work has been supported in part by the EU Contract No. MRTN-CT-2006-035482 (FLAVIAnet).
\end{acknowledgments}
%\mbox{}
%\vskip 1cm

\appendix

\section{Closed formulas for the NRQED o-Ps$\to 3\gamma$ amplitude}
\label{sec:appen1}

We give first some useful formulae for the computations that follow. The partial waves of the Coulomb
Green's function can be written in terms of the associated Laguerre polynomials~\cite{hostler}. Setting one of the arguments
to zero they read,
\begin{eqnarray}
G_\ell \left( 0, x ,\kappa \right) 
&=& 
{m \kappa \over 2 \pi} \left(2 \kappa \right)^{2\ell} e^{-\kappa x } \sum_{n=0}^\infty{ L_n^{2\ell+1}(2 \kappa x)   \over (n+\ell+1-\nu)(2\ell+1)!} \,,
\label{A.1}
\end{eqnarray}
with~\cite{abramowitz}
\begin{eqnarray}
L_n^{2\ell +1}(x)=\sum_{r=0}^{n}\frac{(-1)^r}{r!}{n+2\ell+1 \choose n-r}x^r
\,.
\label{A.2}
\end{eqnarray}
A related integral is
\begin{eqnarray}
J_{n,\ell}^{m}(\alpha)=
\int_{0}^{\infty}dz \, e^{-\alpha z}L_n^{2\ell+1}(z)\,z^m =(-1)^m\frac{d^m}{d\alpha^m}\, J_{n,\ell}^{0}(\alpha)
\,,
\label{A.3}
\end{eqnarray}
that we shall need only for the $\ell=0$ and $\ell=1$ cases:
\begin{eqnarray}
J_{n,0}^{0}(\alpha) &=& -\left( \frac{\alpha-1}{\alpha}\right)^{n+1}+1\nn\\[2mm]
J_{n,1}^{0}(\alpha) &=& -\frac{(\alpha-1)^{n+3}}{\alpha^{n+1}}+ \alpha(\alpha-n-3)+\frac{n^2}{2}+\frac{5}{2}n+3 
\,.
\label{A.4}
\end{eqnarray}
Let us start with the electric amplitude ${\cal D}(\omega)$, Eq.~(\ref{1.14}). Using the representation
(\ref{A.1}) and making the change of variables $z=2\kappa r$, it can be written as:
\begin{eqnarray}
{\cal D}(\omega)&=&16\nu\left( \frac{\kappa}{\omega} \right)^3 \sum_{n=0}^{\infty}\frac{1}{n+2-\nu}
\left\{ i J_{n,1}^0(\alpha)+\frac{\omega}{4\kappa}J_{n,1}^1(\alpha)\right\} +\mbox{h.c.}
\,,
\label{A.5}
\end{eqnarray}
with
$$
\alpha=\frac{1+\nu}{2}-\frac{i\omega}{4\kappa}\,.
$$
(recall from Sec.~\ref{sec:NRQED} that $\nu=1/\kappa a$). With the help of the results (\ref{A.3},\ref{A.4}) and 
the identity,
\begin{eqnarray}
\sum_{n=0}^{\infty}\frac{x^n}{n+a}=\frac{{}_2F_1(a,1;a+1;x)}{a}\,,
\label{A.6}
\end{eqnarray}
we arrive to a closed expression for ${\cal D}(\omega)$ in terms of a hypergeometric function,
\begin{eqnarray}
{\cal D}(\omega) = 16\nu \mu^2 \left( \frac{\kappa}{\omega} \right)^{\! 3 }
\!\bigg\{    {}_2F_1(2-\nu,1;3-\nu;\mu)
\frac{i\alpha(\alpha-1)-\frac{\omega}{4\kappa}(2\alpha+\nu-1)}{\nu-2} 
+\frac{\omega}{4\kappa}\alpha \bigg\}
 +\mbox{h.c.}\nn\\
\label{A.7}
\end{eqnarray}
where we have defined $\mu=(\alpha-1)/\alpha$. The expansion in $(\omega/\kappa)^2$ of the result (\ref{A.7}) is equivalent to the one which is obtained
by directly expanding the trigonometric functions inside the integral representation of ${\cal D}(\omega)$, see Eq.~(\ref{1.19}).
The leading term (dipole approximation) yields a representation of $d_e(\omega)$, Eq.~(\ref{1.21}),
in terms of an
hypergeometric function,
\begin{eqnarray}
d_e(\omega) = -\frac{4}{3}\frac{\nu(1+2\nu)}{(1+\nu)^2} +\frac{8}{3}\,
\frac{\nu^2(\nu-1)}{(\nu-2)(1+\nu)^3} \,{}_2F_1(2-\nu,1;3-\nu;\frac{\nu-1}{\nu+1}) \,.
\label{A.8}
\end{eqnarray}
The representation above agrees with the corresponding one for $a_e(\omega)=1+d_e(\omega)$ found in
\cite{voloshin}. The expansion of $d_e(\omega)$ for small $\nu$ is legitimate
in the $m\alpha^2\ll \omega\ll m$ region, and 
follows immediately from (\ref{A.8}),
\begin{eqnarray}
d_e(\omega) = -\frac{4}{3}\nu+ \frac{8}{3}\left( 1-\log 2 \right)\nu^2 + {\cal O}(\nu^3) \,,
\label{A.9}
\end{eqnarray}
where higher order terms in $\nu$ account for further Coulomb interactions. The subleading
term in the multipole expansion of the electric amplitude corresponds to the $(\omega/\kappa)^2$ term 
in the expansion of
${\cal D}(\omega)$ and so on. Since $(\omega/\kappa)^2\simeq \omega/m$ (see Eq.~(\ref{1.252})), 
multipoles of $r$-th order get suppressed by the overall factor $(\omega/m)^r$.
A formula for the $r$-th multipole is obtained in a more direct way
through the $\omega$-expansion of the trigonometric functions inside ${\cal D}(\omega)$ prior to 
integration. It reads 
\begin{eqnarray}
d_e^{(r)}(\omega) = (-1)^{r+1}\frac{r+1}{(2r+3)!}\left(\frac{\omega}{4\kappa}\right)^{2r}
\sum_{n=0}^\infty \frac{\nu}{n+2-\nu}\,J_{n,1}^{2r+3}(\beta)\,,
\label{A.10}
\end{eqnarray}
with $\beta=(1+\nu)/2$. A closed expression for $d_e^{(r)}(\omega)$ can be obtained with the aid of 
Eqs.~(\ref{A.3},\ref{A.4}) and a generalization of the identity (\ref{A.6}),
\begin{eqnarray}
\sum_{n=0}^{\infty}\frac{n^r}{n+a}x^n=\sum_{\ell=1}^r (-a)^{\ell-1}\left( \mbox{Li}_{\ell-r}(x)+\delta_{r,\ell} \right)
+(-a)^{r-1}\, {}_2F_1(a,1;a+1;x)\,.
\label{A.11}
\end{eqnarray}
For example, the next-to-leading multipole ($r=1$) reads
\begin{eqnarray}
d_e^{(1)}(\omega) &=& \frac{1}{15}\left(\frac{\omega}{\kappa}\right)^2 \! \frac{\nu}{(1+\nu)^4}
\left\{ \frac{\nu^2-3\nu-2}{\nu-1}+\frac{4\nu(1+\nu^2)}{(\nu^2-3\nu+2)(1+\nu)}
\,{}_2F_1(2-\nu,1;3-\nu;\frac{\nu-1}{\nu+1})\right\}\nn \\[2mm]
&=& \frac{2}{15}\nu\left(\frac{\omega}{\kappa}\right)^2  + {\cal O}(\nu^2\omega^2/\kappa^2)
 \,,
\label{A.12}
\end{eqnarray}
whose leading order term in the small $\nu$ expansion agrees with our previous calculation using
the free Green's function in Eq.~(\ref{1.25}). 

Let us also give the exact formula for the full magnetic amplitude ${\cal A}(\omega)$. From
Eq.~(\ref{1.27}), we obtain
\begin{eqnarray}
{\cal A}(\omega)&=& - \frac{im}{2\kappa} \sum_{n=0}^{\infty}\frac{1}{n+1-\nu}
\, J_{n,0}^1(\alpha)+\mbox{h.c.}\nn\\[2mm]
&=& - \frac{im}{2\alpha\kappa} 
\bigg\{ 1+\frac{\nu}{\alpha(1+\nu)} \, {}_2F_1(1-\nu,1;2-\nu;\mu) \bigg\} +\mbox{h.c.}
\,.
\label{A.13}
\end{eqnarray}
As in the case of the electric amplitude, the multipole expansion of ${\cal A}(\omega)$
is realized through a series in $(\omega/\kappa)^2$. The exact result for the magnetic
dipole term, $a_m(\omega)$,
has been already shown in Eq.~(\ref{1.29}). The next-to-leading term in the multipole expansion reads
\begin{eqnarray}
a_m^{(1)}(\omega) &=& -\frac{1}{12}\frac{m\omega}{\kappa^2}\,\left(\frac{\omega}{\kappa}\right)^2 \frac{1}{(1-\nu^2)^3(1+\nu)}
\nn\\[2mm]
&&\times\left\{ 3-\nu+\nu^2+5\nu^3 -8\nu^2\,
\,{}_2F_1(1-\nu,1;2-\nu;\frac{\nu-1}{\nu+1})\right\}\,.
\label{A.14}
\end{eqnarray}
The expansion of $a_m^{(1)}(\omega)$ for small $\nu$ gives
\begin{eqnarray}
a_m^{(1)}(\omega)
&=& -\frac{\omega}{4m}\left(1+\frac{\omega}{4m}\right)^{-2}\,\Big(1-\frac{4}{3}\nu+ {\cal O}(\nu^2)\Big) 
 \,,
\label{A.15}
\end{eqnarray}
where we have made use of the relation between $\kappa^2$ and $\omega$ given in Eq.~(\ref{1.252}). Finally, we
give the general form of a the $r$-th order magnetic multipole:
\begin{eqnarray}
a_m^{(r)}(\omega) = \frac{(-1)^{r}}{(2r+1)!}\,\frac{m\omega}{4\kappa^2}\left(\frac{\omega}{4\kappa}\right)^{2r}
\sum_{n=0}^\infty \frac{1}{n+1-\nu}\,J_{n,0}^{2r+2}(\beta)\,,
\label{A.16}
\end{eqnarray}
which scales as $(\omega/m)^r$.

\section{Threshold expansion of a triangle diagram with a soft radiated photon}
\label{sec:appen2}

In this appendix we show a further example of the application of the asymptotic expansion method 
near threshold to a 1-loop diagram with massive lines 
and a soft momentum component radiated off. The diagram is shown in Fig.~\ref{fig:threshold2}.
This example is similar to the 1-loop ladder diagram considered in 
Sec.~\ref{sec:threshold}, but we shall not take here the total incoming momentum $P^2$
equal to $4m^2$, 
but instead define $y=m^2-P^2/4\sim (m\alpha)^2$. The parameter $y$, which 
appears in the massive propagators, will give rise to a non-zero contribution from the potential 
region, which was not present in the ladder diagram evaluated at threshold.
For simplicity, we have substituted the Coulomb-ladder interaction by
an effective production vertex. We shall take the soft photon momentum $k\sim m\alpha$,
and keep terms in the asymptotic expansion up ${\cal O}(\alpha)$.

If we consider scalar propagators, the integral of Fig.~\ref{fig:threshold2} is given by
\begin{eqnarray}
I_2 
& =&  \int
{  [d^Dq]\over  (q^2+q\cdot P-y) \, (q^2-q\cdot P-y) \,  (q^2+q\cdot P-2q\cdot k - P\cdot k-y) }
\,,
\label{B.1}
\end{eqnarray}
where we have choosen a routing of the external momentum $P$ through the massive lines of the graph
which allows to use the scaling arguments defined in Eqs.~{(\ref{3.1})} and~{(\ref{3.2})}. When the loop
momentum is hard ($q\sim m$), we can expand the propagators in the small variables $y$ and $k$. 
Keeping terms up to ${\cal O}(\alpha)$ the expansion
in the hard region reads:
\begin{eqnarray}
I_2^{(\text{h})} 
& =&  \int
{  [d^Dq]\over  (q^2+q\cdot P)^2 \, (q^2-q\cdot P)}
\left(1+{2 q\cdot k + P\cdot k\over q^2+q\cdot P}+\dots \right)\nn\\[2mm]
&=&\frac{1}{32m^2\pi^2 } \left( 1+\frac{\omega}{3m}\right)+{\cal{O}}(\alpha^2)\,,
\label{B.2}
\end{eqnarray}
and higher order terms can be calculated straightforwardly. 
The first region that we shall consider when the loop momentum is small is the soft-radiation region, that
we already found in the 1-loop diagrams contributing to the o-Ps$\to 3\gamma$ amplitude.
According to the hierarchy $q^0\sim \bmq^2/m \sim \omega \sim m\alpha$ we expand the propagators retaining terms up to
${\cal O}(\alpha)$:
%
%%%%%%%%%%%%%%%%%%%%%%%%%%%%%%%%%%
\begin{figure}
\begin{center}
\includegraphics[width=5cm]{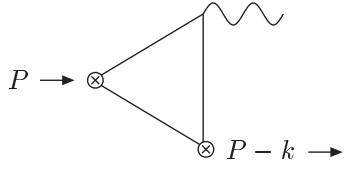}
\end{center}
\caption{Scalar triangle diagram. \label{fig:threshold2}}
\end{figure} 
%%%%%%%%%%%%%%%%%%%%%%%%%%%%%%%%%%
\begin{eqnarray}
I_2^{(\text{sr})} %& =&  \int
%{  [d^Dq]\over   [(q+P/2)-m^2] \, [(q-P/2)^2-m^2] \,  [(q+P/2-k)^2-m^2] }\\
& =&  \int
{  [d^Dq]\over  (-\bmq^2+q_0 P_0 -y) \, (-\bmq^2-q_0 P_0 -y)\, (-\bmq^2+q_0 P_0-\omega P_0 )}\nn\\[2mm]
&& \times
\left(1+{ y-q_{0}^2 \over -\bmq^2+q_0 P_0  -y}+{ y-q_{0}^2 \over -\bmq^2-q_0 P_0 -y}
+{ y+2q_{0}\omega-q_0^2 \over -\bmq^2+q_0 P_0- \omega P_0} \right. \nn\\[2mm]
&& \left. \quad\;\; + \,  { 4(\bmq\cdot \mathbf{k})^2 \over (-\bmq^2+q_0 P_0- \omega P_0  )^2}+
\dots \right)\nn\\[2mm]
&=&-\frac{1}{32m^2\pi} \sqrt{\frac{m}{\omega}} \left( 1+ \frac{y}{2m\omega}+\frac{\omega}{6m} \right)
+{\cal{O}}(\alpha^{3/2})\,.
\label{B.3}
\end{eqnarray}
The integrals that result from Eq.~(\ref{B.3}) are calculated easily with standard methods. The leading term
in the soft-radiation region is enhanced by $1/\sqrt{\alpha}$ with respect the hard region.
The potential region, where $q^0\sim \bmq^2/m \sim y/m$, also contributes when the loop momentum is small.
The expansion of the integrand gives:
\begin{eqnarray}
I_2^{(\text{p})} %& =&  \int
%{  [d^Dq]\over   [(q+P/2)-m^2] \, [(q-P/2)^2-m^2] \,  [(q+P/2-k)^2-m^2] }\\
& =& -\frac{1}{P\cdot k} \int
{  [d^Dq]\over  (-\bmq^2+q_0 P_0 -y) \, (-\bmq^2-q_0 P_0-y)}
\left(1+{ -\bmq^2+q_0P_0-y \over P\cdot k}+\dots \right)\nn\\[2mm]
&=&\frac{1}{32m^2\pi} \frac{\sqrt{y}}{\omega}+{\cal{O}}(\alpha^2)\,.
\label{B.4}
\end{eqnarray}
The term $P\cdot k$ dominates the massive propagator depending on $k$, which shrinks to a point
in the potential region.
The subleading term between parentheses in Eq.~(\ref{B.4}) arises from the expansion of the latter
propagator, but gives a vanishing contribution because the numerator is proportional to one of the 
propagators in front. Subleading terms in the expansion of the propagators not depending on $k$ generate
corrections which scale as $(q_0)^2/(-\bmq^2+q_0P_0-y )\sim \alpha^2$, that we do not retain. 

The soft loop-momentum region leads to a scaleless $\bmq$-integration after
the $\bmq$ dependence has been expanded out from the denominators according to 
the scaling $q_0\sim \bmq \sim m\alpha$:
\begin{eqnarray}
I_2^{(\text{s})} 
& =&  \int
{  [d^Dq]\over  (q_0 P_0) \, (-q_0 P_0)\, (q_0 P_0-\omega P_0)} +\dots =0\,,
\label{B.5}
\end{eqnarray}
and the same holds for the ultrasoft region.
Contributions from the soft and ultrasoft regions can arise in diagrams with
massless propagators, like the 1-loop ladder graph of Sec.~\ref{sec:threshold}. Apart
from the latter, the analysis of the threshold expansion of the 1-loop ladder diagram for $P^2\ne 0$ only
leads to trivial modifications to the integrations for the various regions found here,
and can be obtained easily.

Finally, the exact result for $I_2$ when $k^2=0$ 
can be read off the list of integrals in Ref.~\cite{adkins}:
\begin{eqnarray}
I_2 & =& 
 -\frac{1}{8\pi^2}\frac{1}{P\cdot k}
\left\{ \mbox{L}\left(\frac{(P-k)^2}{m^2}\right) -\mbox{L}\left(\frac{P^2}{m^2}\right) \right\}
\,,
\label{B.6}
\end{eqnarray}
with
\begin{eqnarray}
\mbox{L}(s) & =& -2\left( \mathrm{arctan}\sqrt{\frac{s}{4-s}} \,\right)^2
\,.
\label{B.7}
\end{eqnarray}
One can verify that the expansion of the exact result for $\omega\sim m\alpha$ up to terms
of ${\cal O}(\alpha)$ is reproduced by the sum of the contributions of the regions in 
(\ref{B.2}),(\ref{B.3}) and (\ref{B.4}).

%
%Bibliography

\end{document}